%% file: lux_run03_dd_prc.tex
\documentclass[aps,prc,reprint,twocolumn,superscriptaddress,nofootinbib,floatfix]{revtex4-1}

\usepackage{amsmath}
\usepackage{amssymb}
\usepackage{graphicx}
\usepackage{enumerate}
\usepackage{siunitx}
\usepackage{dcolumn}%
\usepackage{bm}%
\usepackage{color}
\definecolor{navyblue}{rgb}{0.0, 0.0, 0.5}

\usepackage[pdftex,colorlinks=true,citecolor=navyblue,filecolor=navyblue,linkcolor=navyblue,urlcolor=navyblue]{hyperref}

\graphicspath{ {./figures/} }

\newcommand{\kevnr}{\ensuremath{\textrm{keV}_\textrm{nr}}} %
\newcommand{\kevee}{\ensuremath{\textrm{keV}_{\textrm{ee}}}} %

\newcommand{\leff}{$\mathcal{L}_{\textrm{eff}}$} %
\newcommand{\qy}{$Q_{y}$} %
\newcommand{\ly}{$L_{y}$} %
\newcommand{\dd}{$\mbox{D-D}$} %

\newcommand{\np}{$n_{p}$}%
\newcommand{\nel}{$n_{e}$} %
\newcommand{\nq}{$n_{q}$}

\newcommand{\Nph}{$N_{\textrm{ph}}$}
\newcommand{\Ne}{$N_{e}$}

\newcommand{\se}{$s_{e}$}
\newcommand{\sn}{$s_{n}$}

\newcommand{\insitu}{$\textit{in situ}$}
\newcommand{\exsitu}{$\textit{ex situ}$}

\begin{document}

\title{Low-energy (0.7--74~keV) nuclear recoil calibration of the {LUX} dark matter experiment using {D}-{D} neutron scattering kinematics}

\input{lux_run03_dd_author_list.tex}

\collaboration{LUX Collaboration}\noaffiliation 

\date{\today}

\begin{abstract}
    \noindent
    The Large Underground Xenon (LUX) experiment is a dual-phase liquid xenon time projection chamber (TPC) operating at the Sanford Underground Research Facility in Lead, South Dakota.
    A calibration of nuclear recoils in liquid xenon was performed \insitu{} in the LUX detector using a collimated beam of mono-energetic 2.45 MeV neutrons produced by a deuterium-deuterium (\dd{}) fusion source.
    The nuclear recoil energy from the first neutron scatter in the TPC was reconstructed using the measured scattering angle defined by double-scatter neutron events within the active xenon volume. 
    We measured the absolute charge (\qy{}) and light (\ly{}) yields at an average electric field of 180~V/cm for nuclear recoil energies spanning 0.7 to 74~keV and 1.1 to 74~keV, respectively.
    This calibration of the nuclear recoil signal yields will permit the further refinement of liquid xenon nuclear recoil signal models and, importantly for dark matter searches, clearly demonstrates measured ionization and scintillation signals in this medium at recoil energies down to $\mathcal{O}$(1 keV).
\end{abstract}

\maketitle

\section{Introduction}

The Large Underground Xenon (LUX) detector is a 370~kg (250~kg active mass) dual-phase liquid xenon TPC designed to directly detect WIMP dark matter in the local galactic halo.
The detector is located at the center of an 8~m diameter, 6~m tall water shield on the 4850$^{\prime}$ level of the Sanford Underground Research Facility (SURF).
The monolithic liquid xenon target is instrumented with 61 photomultiplier tubes (PMTs) in a top array and 61 PMTs in a bottom array.
A detailed description of the LUX detector design is available in Ref.~\cite{AkeribBaiBedikianEtAl2013}.
Recent WIMP search results from the LUX detector have placed the most stringent direct detection limits on the spin-independent WIMP-nucleon cross-section~\cite{Akerib2014, AkeribAraujoBaiEtAl2015} and the WIMP-neutron spin-dependent cross-section~\cite{AkeribAraujoBaiEtAl2016} for a large range of WIMP masses.
Establishing the sensitivity of the LUX detector to nuclear recoil events arising from particle interactions requires an accurate calibration of the signal response of liquid xenon over the expected range of energy transfers.

Energy depositions in the liquid xenon target produce both scintillation photons and ionization electrons.
The prompt scintillation photon signal (S1) is directly detected by the PMTs. 
The ionization electrons drift at a constant average drift speed to the liquid surface under an average applied electric field of 180~V/cm, where they are extracted into the gas region and produce a signal (S2) in the PMTs via photon emission due to electroluminescence.
Ionization electrons can attach to electronegative impurities in the liquid xenon---an effect that exponentially suppresses the observed S2 signal size~\cite{Aprile2010a}. 
The characteristic time constant associated with this process is referred to as the electron lifetime.
The pulse areas associated with the S1 and S2 signals are position-corrected as described in Ref.~\cite{AkeribAraujoBaiEtAl2015, Akeribothers2016} and are referred to using the variables $S1$ and $S2$ (note the italics when referring to the measured quantity). 
The signal corrections are described in more detail in Sec.~\ref{sec:lux_dd_detector_operating_parameters}.
In the several instances where uncorrected S1 and S2 signal sizes are used, the variables will be labeled as ``raw'' $S1$ and $S2$.
The raw and corrected variables $S1$ and $S2$ are given in units of detected photons (phd).
The unit ``phd'' differs from the traditional unit of photoelectrons (phe) by accounting for the probability of double photoelectron emission from a single absorbed VUV photon~\cite{FahamGehmanCurrieEtAl2015}, which was measured for each LUX PMT.
The nuclear recoil band analysis described in Sec.~\ref{sec:dd_nr_band} uses an alternative technique to characterize the S1 signal size.
When using this technique, the S1 signal size, represented by the variable $S1_{\textrm{spike}}$, is measured by counting the number of single photon ``spikes'' in the per-channel waveforms.
This is the same S1 signal size variable used in the recent WIMP search results~\cite{AkeribAraujoBaiEtAl2015, AkeribAraujoBaiEtAl2016}.

The single quanta gain values for scintillation photons ($g_1$) and ionization electrons ($g_2$) escaping the particle interaction site were calibrated directly in LUX and have units of phd~per~scintillation~photon and phd~per~ionization~electron, respectively~\cite{AkeribAraujoBaiEtAl2015, Akeribothers2016}. 
The value $g_1$ is dictated by the product of the fraction of scintillation light collected and the detection efficiency of the PMTs. 
The value $g_2$ is dictated by the extraction efficiency of electrons from the liquid and the electroluminescent yield in the gas.
This analysis uses $g_1$ of 0.115~$\pm$~0.004 and a $g_2$ of 11.5~$\pm$~0.9.
These gain values were adjusted from the WIMP search values for the nuclear recoil calibration time period as described in Sec.~\ref{sec:lux_dd_detector_operating_parameters}.
They are within 1.7\% and 4.6\% of the WIMP search values, respectively~\cite{AkeribAraujoBaiEtAl2015}.
These precisely measured $g_1$ and $g_2$ values allow us to directly report the nuclear recoil signal yields in liquid xenon in terms of the absolute number of quanta produced.
This is particularly notable in the case of the light yield result, which is the first direct measurement of nuclear recoil scintillation reported in absolute units.

\subsection{Discussion of historical notation for nuclear recoil signal yields}

It is fairly straightforward to measure the electron recoil signal yields in units of quanta per unit energy using line sources that fully deposit their known energy in the liquid xenon at a single site~\cite{ManalaysayUndagoitiaAskinEtAl2010, Akeribothers2016}.
Absolute calibration of the detector's response to nuclear recoils induced via neutron scattering is more challenging for several reasons:
unlike electron recoil calibrations, there are no convenient sealed or injectable sources providing mono-energetic neutrons;
due to the variable energy transfer to the nucleus depending upon the scattering angle, even mono-energetic neutrons produce a range of recoil energies;
only a small fraction of the incident neutron energy is deposited at each interaction, and the neutron mean free path of $\mathcal{O}$(10~cm) typically results in multiple-site interactions in the detector medium and energy-loss in passive detector materials.
To unambiguously identify the source of energy depositions in the liquid xenon, we use the units \kevee{} and \kevnr{} for electron and nuclear recoils, respectively.

The $S1$ for liquid xenon TPCs is traditionally related to the nuclear recoil energy deposited at the interaction site, $E_{\textrm{nr}}$ via

\begin{equation} \label{eq:recoil_energy_to_s1_leff}
    S1 = E_{\textrm{nr}} ~ \mathcal{L}_{\textrm{eff}}(E_{\textrm{nr}}) ~ L_{y,\,^{57}\textrm{Co}}(\mathcal{E}) ~ \frac{S_{\textrm{nr}}(\mathcal{E})}{S_{\textrm{ee}}(\mathcal{E})} \,\text{,}
\end{equation}

\noindent
where \leff{}($E_{\textrm{nr}}$) is the scintillation yield for nuclear recoils relative to the scintillation signal produced by the 122~\kevee{} $^{57}$Co gamma ray at 0~V/cm~\cite{Manzur2010}.
When operating at non-zero drift electric field $\mathcal{E}$, the scintillation signal from both nuclear and electron recoil interactions is quenched to a fraction of the 0~V/cm value~\cite{Aprile2006a}.
The quenching fractions for nuclear and electron recoil signals are represented by $S_{\textrm{nr}}$ and $S_{\textrm{ee}}$, respectively.
The measured light yield for electron recoils from the $^{57}$Co gamma ray at the TPC operating drift field is represented by $L_{y,\,^{57}\textrm{Co}}(\mathcal{E})$.
The quantity $L_{y,\,^{57}\textrm{Co}}(\mathcal{E})$ is detector-dependent and includes effects such as the photon detection efficiency.
The 122~\kevee{} $^{57}$Co gamma ray has an attenuation length of 2~mm in liquid xenon, which is not well suited for calibration in large TPCs such as LUX (0.5~m linear dimension).

The $S2$ is related to the recoil energy deposited at the interaction site by

\begin{equation} \label{eq:recoil_energy_to_s2_qy}
    S2 = E_{\textrm{nr}} ~ Q_{y}(E_{\textrm{nr}}, \mathcal{E}) ~ g_2 \,\text{,}
\end{equation}

\noindent
where \qy{}($E_{\textrm{nr}}$, $\mathcal{E}$) is the ionization yield for nuclear recoils at the applied electric field given in the absolute units of electrons/\kevnr{}.

Numerous measurements of both \leff{} and \qy{} at low energies exist in the literature, primarily motivated by the need to understand and calibrate the liquid xenon signal response for WIMP dark matter searches.
Various experimental strategies are used to measure these quantities:
\begin{enumerate}[i.]
    \item Nuclear recoil calibrations performed \insitu{} in the dark matter detector itself via simulation-based best-fit models optimized to match the observed signal spectrum from neutron sources with a continuous energy spectrum~\cite{Sorensen2010a, Horn2011, AprileAlfonsiArisakaEtAl2013}.
    \item Mono-energetic neutron \exsitu{} calibrations in a small liquid xenon test cell using the neutron scattering angle to kinematically define the recoil energy~\cite{AprileBaudisChoiEtAl2009, Manzur2010, Plante2011}.
    \item Nuclear recoil spectrum endpoint based calibrations~\cite{JoshiSangiorgioBernsteinEtAl2014}.
\end{enumerate}
The advantages and disadvantages of the various techniques are discussed in Refs.~\cite{Verbus2016, VerbusRhyneMallingEtAl2016}.

\subsection{Discussion of modern notation for nuclear recoil signal yields}

Here, high-precision measurement of $g_1$ allows the scintillation yield result to be reported absolutely~\cite{Akeribothers2016}.
The energy calibration of the $S1$ detector response is expressed directly in terms of the light yield for nuclear recoils at electric field $\mathcal{E}$ in units of photons/\kevnr{}, \ly{}($E_{\textrm{nr}}$, $\mathcal{E}$), as: 

\begin{equation} \label{eq:recoil_energy_to_s1_ly}
    S1 = E_{\textrm{nr}} ~ L_{y}(E_{\textrm{nr}}, \mathcal{E}) ~ g_1 \,\text{.}
\end{equation}

\noindent
Reporting the light yield in absolute terms at the operating electric field as described in Eq.~\ref{eq:recoil_energy_to_s1_ly} also has the advantage of avoiding any assumptions about the field quenching factors ($S_{\textrm{nr}}$ and $S_{\textrm{ee}}$).

\subsection{Results reported in this article}

For the results presented here, a deuterium-deuterium (\dd{}) neutron generator was used as the neutron source. 
We use event-by-event kinematic reconstruction of neutron double-scatters in the TPC to obtain an absolute measurement of the nuclear recoil energy $E_{\textrm{nr}}$, and combine this with the LUX absolute calibration of the LUX ionization channel (using $g_2$) to obtain a direct calibration of \qy{}~\cite{Gaitskell2015}.
Following Eq.~\ref{eq:recoil_energy_to_s2_qy}, this \qy{} measurement provides a precise calibration of $S2$ as a function of recoil energy, which is used to extract \ly{} from the single-scatter event population using $g_1$ to determine the absolute number of S1 photons collected per Eq.~\ref{eq:recoil_energy_to_s1_ly} (further discussion in Sec.~\ref{sec:dd_low_energy_ly_data_analysis}).
Additionally, we use the known nuclear recoil energy spectrum endpoint for neutrons produced by our \dd{} source, again combined with the LUX $g_1$ and $g_2$, to absolutely measure \qy{} and \ly{} at 74~\kevnr{}. 
By determining the yields \insitu{} in the dark matter instrument itself, one avoids potential systematic uncertainties intrinsic in the translation of \exsitu{} measurements.

This new nuclear recoil calibration refines the LUX nuclear recoil signal detection efficiency estimates and also proves the kinematic accessibility of more WIMP-mass parameter space, given the local galactic escape and Earth-halo velocities.
Prior to this result, the lowest-energy light and charge yield results determined using a kinematically-defined nuclear recoil energy scale were reported at 3~\kevnr{}~\cite{Plante2011} and 4~\kevnr{}~\cite{Manzur2010}, respectively.
Due to the absence of any nuclear recoil calibrations in the literature for kinematically-defined nuclear recoil energies $<$3~\kevnr{}, the first LUX spin-independent WIMP search sensitivity result was conservatively limited by assuming no signal yield $<$3~\kevnr{}, where detector efficiency was nevertheless expected to be significant (and, in retrospect, was)~\cite{Akerib2014}.
This new LUX calibration result documents an improvement in the instrument's sensitivity at low WIMP masses using the existing 2013 WIMP search dataset by demonstrating signal yield in both channels for nuclear recoil energies as low as 1.1~\kevnr{}~\cite{AkeribAraujoBaiEtAl2015}.

This paper is organized as follows.
The experimental setup for the LUX neutron calibration is described in Sec.~\ref{sec:lux_dd_experimental_setup}.
A low-energy (0.7--24.2~\kevnr{}) measurement of \qy{} using the measured scattering angle between double-scatter event interaction sites in the TPC is presented in Sec.~\ref{sec:dd_low_energy_qy}.
A low-energy (1.1--12.8~\kevnr{}) measurement of \ly{} using single-scatter events is reported in Sec.~\ref{sec:dd_low_energy_ly}.
The \qy{} and \ly{} at the 74~\kevnr{} recoil energy spectrum endpoint is reported in Sec.~\ref{sec:dd_endpoint}.
The LUX nuclear recoil band measurement is shown in Sec.~\ref{sec:dd_nr_band}.
A different set of event selection cuts is appropriate for each of these analyses.

The specific cuts used for each analysis are outlined at the beginning of each section.
Two new NEST~\cite{LenardoKazkazManalaysayEtAl2015} nuclear recoil models (one based on the Lindhard model~\cite{Lindhard1963a}, one based on the Bezrukov parameterization~\cite{Bezrukov2011}) were created via a simultaneous fit to all \qy{}, \ly{}, and nuclear recoil band results reported in this article.
These new NEST models are described in Sec.~\ref{sec:nest_post_dd}.

The results presented in this paper used several simulation frameworks to produce targeted results as appropriate for each section.
The Monte Carlo setup used for each section is described in the text.

\section{Experimental setup} \label{sec:lux_dd_experimental_setup}

\begin{figure}[!htbp]
    \begin{center}
        \includegraphics[width=0.480\textwidth]{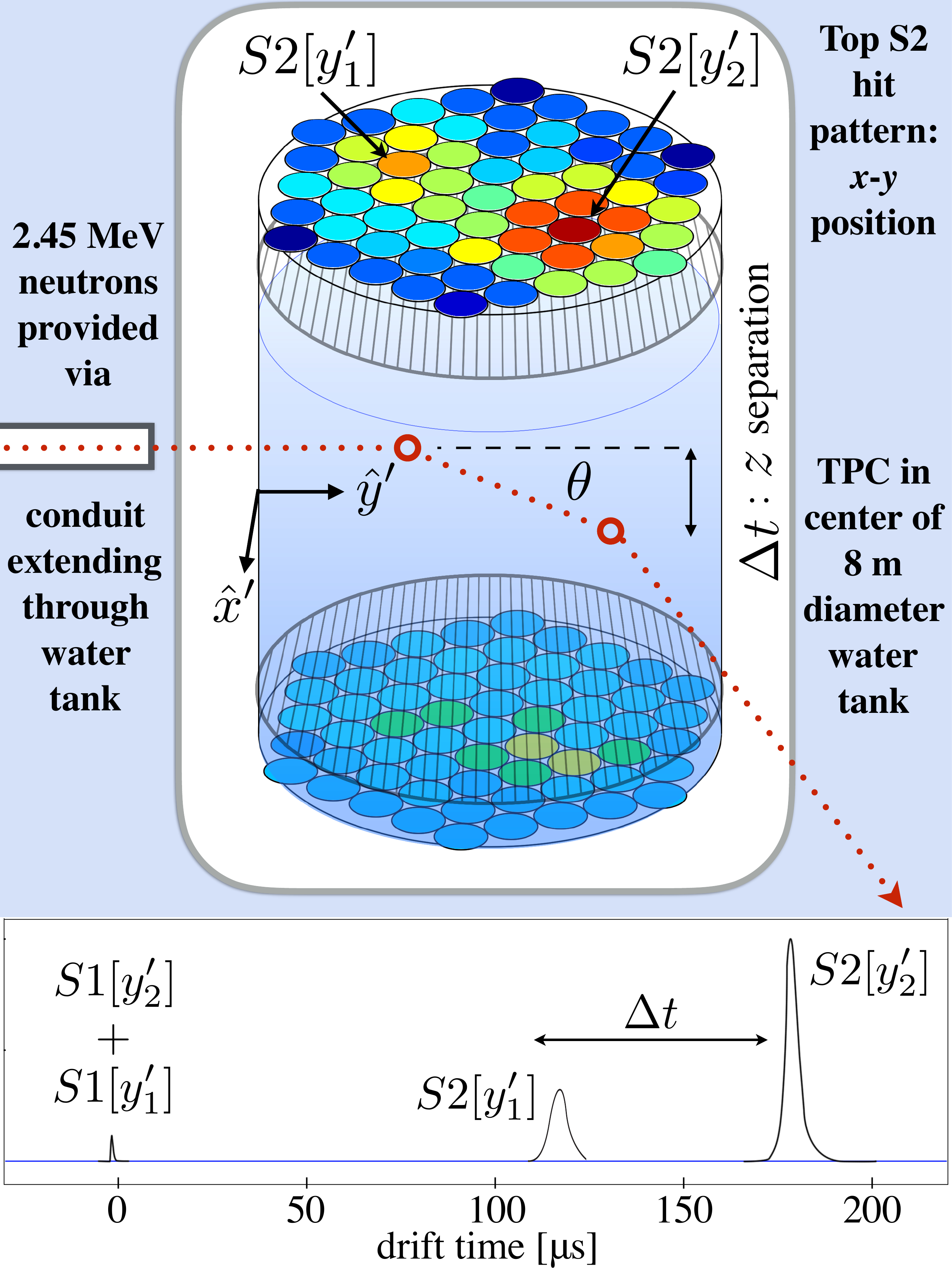}
        \vskip -0.1cm
        \caption{
            Conceptual diagram of the LUX \dd{} calibration experimental setup.
            The LUX TPC is in the center of the 8~m diameter, 6~m tall water shield.
            The LUX cryostat boundary is depicted as the thick gray line around the TPC.
            The TPC active region has a diameter of 47~cm and a height of 48~cm~\cite{Akerib2014}.
            The mono-energetic 2.45~MeV neutrons are collimated through an air-filled conduit spanning the distance from the water tank wall to the LUX detector cryostat.
            The $x^{\prime}$ coordinate is coming out of the paper, and the $y^{\prime}$ coordinate is in line with the beam.
            This figure illustrates a potential event used for the \qy{} analysis: a neutron (red dotted line) enters the active liquid xenon volume, scatters twice, and then leaves the target media.
            The resulting time-integrated hit pattern is shown on the PMT arrays.
            The bottom frame shows an event record of this neutron interaction sequence (for illustration only).
            The PMT hit pattern provides ($x$,\,$y$) information, while the electron drift time to the liquid surface provides precise reconstruction of the $z$ position of each neutron interaction.
        } 
        \vskip -0.5cm
        \label{fig:lux_dd_scatter_diagram}
    \end{center}
\end{figure}

The experimental setup at the TPC is shown in Fig.~\ref{fig:lux_dd_scatter_diagram}.
Neutrons produced by the \dd{} source are introduced into the TPC via an air-filled conduit spanning the LUX water tank as described in Sec.~\ref{sec:the_neutron_beam}.
A convenient coordinate system used for the subsequent nuclear recoil calibration analysis is defined here.
The orientation of the Cartesian coordinates $x^{\prime}$, $y^{\prime}$, $z^{\prime}$ are defined by the neutron beam pipe axis.
The neutron beam spans a geometrical chord that is offset from the TPC diameter.
The coordinate $y^{\prime}$ is along the beam pipe direction with zero at the point where the beam enters the liquid xenon active region.
The coordinate $x^{\prime}$ is transverse to the beam pipe axis in the horizontal plane.
The $x^{\prime}$ and $y^{\prime}$ coordinates defined by the beam direction differ from the more traditional $x$ and $y$ coordinates, which are centered in the middle of the TPC, by the translation and rotation defined by

\begin{equation} \label{eq:dd_prime_coordinate_rotation}
    \begin{bmatrix} x^{\prime} \\ y^{\prime} \end{bmatrix} =
    \begin{bmatrix} \cos{\theta_{\textrm{rot}}} & -\sin{\theta_{\textrm{rot}}} \\ \sin{\theta_{\textrm{rot}}} & \cos{\theta_{\textrm{rot}}} \end{bmatrix}
    \begin{bmatrix} x - 7.1~\textrm{cm} \\ y + 23.0~\textrm{cm} \end{bmatrix}
    \,\text{,}
\end{equation}

\noindent
where $\theta_{\textrm{rot}} = -5.1^{\circ}$.
The coordinate $z^{\prime}$ is perpendicular to the beam pipe axis in the vertical plane.
It is nearly identical to the traditional $z$ (ionization drift) coordinate indicating the distance from the liquid surface.
The neutron beam entry point into the liquid xenon volume is 0.9~cm above the exit point.
This corresponds to an angle of ${\sim}1^{\circ}$ with respect to the liquid xenon surface.\footnote{The small angle of the neutron conduit with respect to the liquid xenon surface was due to the precision of the neutron conduit leveling process.}
A distance of 47.4~cm along $y^{\prime}$ separates the entry and exit points of the neutron beam in the liquid xenon.

This notation is further used in this paper such that $S2[y_{1}^{\prime}]$ and $S2[y_{2}^{\prime}]$ represent the S2 signal size from the first and second neutron-xenon scattering sites in the $y^{\prime}$ direction along the beam line, respectively.
The $z$~vs.~$y^{\prime}$ distribution of single-scatter events is shown in Fig.~\ref{fig:xyz_neutron_distribution_z_vs_yprime_dist_single_scatters_densityplot_with_beam_purity_fiducial}.

We use the coordinate $r$, which is the radial coordinate in the cylindrical coordinate system coaxial with the monolithic liquid xenon target. 

\subsection{LUX detector operating parameters} \label{sec:lux_dd_detector_operating_parameters}

The nuclear recoil calibration program using a \dd{} neutron generator discussed in this paper was performed at the end of the 2013 LUX WIMP search run using the same detector operational state, including identical DAQ/trigger conditions and frequent $^{83\textrm{m}}$Kr-based calibrations for position-dependent S1 and S2 signal corrections~\cite{AkeribAraujoBaiEtAl2015}.
As in Ref.~\cite{AkeribAraujoBaiEtAl2015}, the event window extends $\pm$500~$\mu$s around the trigger signal generated by the hardware trigger. 
For S2 signals produced by nuclear recoils in the beam line, the mean electron lifetime correction was $1.16 \times S2$ and the average ($x$,\,$y$) correction was $0.96 \times S2$.
For S1 signals produced by nuclear recoils in the beam line, the mean ($x$,\,$y$,\,$z$) position correction was $1.06 \times S1$.
Data were corrected for any time variation between their direct measurement during the WIMP search period and the later \dd{} calibration period using the variation in $^{83\textrm{m}}$Kr $S1$ and $S2$ peak positions.
The single electron (SE) distribution was measured to have a mean value of $23.77 \pm 0.01$~phd during the \dd{} measurements with a standard deviation of $5.75 \pm 0.01$~phd.
The electron extraction efficiency during the \dd{} calibration period was $0.48 \pm 0.04$.
The average electron drift velocity was measured to be $1.51 \pm 0.01$~mm/$\mu$s corresponding to a 324~$\mu$s maximum drift time~\cite{Akerib2014}.

The systematic uncertainty in $S1$ and $S2$ due to time variation in the three-dimensional (3D) position-based corrections using $^{83\textrm{m}}$Kr was determined to be 0.6\% and 2.5\%, respectively.
A small radial drift field component alters the path of drifting electrons in the liquid xenon, with a maximum inward radial deflection of 4.6~cm for electrons originating at the bottom of the TPC~\cite{AkeribAraujoBaiEtAl2015}.
The magnitude of this radial component is smaller near the liquid surface where the neutron beam is positioned.
The reconstructed event position is corrected to account for this effect. 
The systematic uncertainty in $S1$ and $S2$ from the $^{83\textrm{m}}$Kr-based corrections due to these non-uniformities in the drift field was determined to be a bias of 0.5\% in $S1$ and 2.5\% in $S2$.

\subsection{The neutron beam} \label{sec:the_neutron_beam}

\begin{figure}[!htbp]
    \begin{center}
        \includegraphics[width=0.48\textwidth]{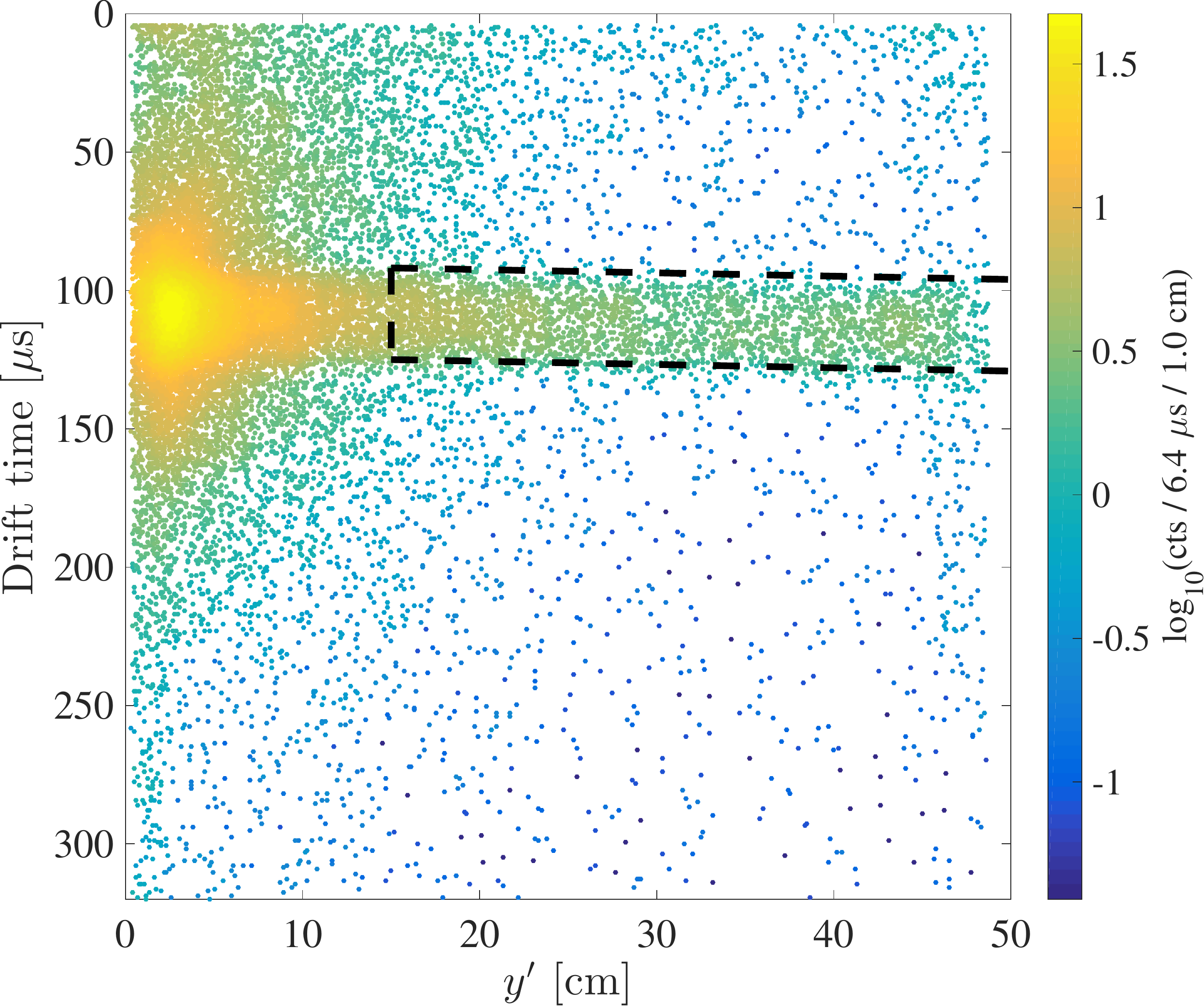}
        \vskip -0.1cm
        \caption{
            The $z$ (drift time) vs. $y^{\prime}$ distribution of single-scatter events passing all nuclear recoil area selection and data quality cuts.
            The neutron beam pipe is aligned outside of the plot to the left in line with the beam at a drift time of 107~$\mu$s.
            A position cut was used to select scatters in a 10~cm wide slide in $x^{\prime}$ around the projection of the neutron beam into the liquid xenon.
            This plot contains the full 107.2~live-hours of 2013 \dd{} data.
            The shine due to neutrons scattering in passive detector materials can be seen localized where the beam enters the liquid xenon.
            The black dashed line shows the approximate location of the neutron beam energy purity cuts.
            The neutron shine near the beam entry point is asymmetric in this plot due to the event selection criteria; only single-scatter events are accepted for this plot and the 12.6~cm total mean free path for neutrons makes it more probable for a neutron to exit out of the top of the xenon volume than the bottom.
        }
        \vskip -0.5cm
        \label{fig:xyz_neutron_distribution_z_vs_yprime_dist_single_scatters_densityplot_with_beam_purity_fiducial}
    \end{center}
\end{figure}

An Adelphi Technology, Inc. DD108 neutron generator was used as the mono-energetic neutron source. 
The neutron generator was operated externally to the LUX water tank shield.
Neutrons were introduced into the LUX detector via a narrow air-filled pipe, which displaced water producing a collimation path.
The sealed, air-filled 4.9~cm inner-diameter (ID), 6.0~cm outer-diameter (OD) polyvinyl chloride (PVC) conduit was suspended by stainless steel wire rope from the top of the 6~m tall LUX water tank.
The neutron conduit is 377~cm in length, spanning the horizontal distance from the outer water tank wall to the outer surface of the LUX cryostat.
The sum of the water gaps at the two ends of the conduit is 6~cm.
During the nuclear recoil calibration campaign, the center of the neutron conduit was raised to be 16.1~cm below the liquid xenon surface in the TPC and leveled to 1$^{\circ}$ with respect to the liquid surface as shown in Fig.~\ref{fig:xyz_neutron_distribution_z_vs_yprime_dist_single_scatters_densityplot_with_beam_purity_fiducial}.
This $z$ position of the beam was chosen to provide a short distance to the liquid surface in order to increase the fraction of low-multiplicity neutron scatters in the dataset.
The observed profile of single-scatter neutron events was used to define the direction of the neutron beam through the TPC.
The shape of the observed beam profile is consistent with the expectation from the solid angle presented by the neutron calibration conduit.
The source of neutron production inside the neutron generator was positioned $46 \pm 2$~cm outside of the LUX water tank in line with the neutron conduit during the calibration.
The neutron conduit was stored out of line with the TPC during the WIMP search campaign.

The energy spectrum of the specific DD108 hardware was characterized at Brown University prior to use in the LUX calibration~\cite{VerbusRhyneMallingEtAl2016}.\footnote{The LUX calibration used an identical shielding structure and source configuration defined as ``Target Orientation A'' in Ref.~\cite{VerbusRhyneMallingEtAl2016}.}
The mean neutron energy was measured to be $2.40 \pm 0.06$~MeV, consistent with the expected 2.45~MeV. 
The expected mean neutron energy of 2.45~MeV was used for the LUX nuclear recoil signal yield data analysis with an associated uncertainty of 2\%.
For the LUX calibration, the DD108 source was operated at 5\% duty cycle using 100~$\mu$s neutron pulses at a 500~Hz repetition rate.
An incident neutron flux of $78 \pm 8$~n\,cm$^{-2}$\,s$^{-1}$ was measured on the exterior of the water tank near the entrance to the calibration conduit using a 9~inch diameter Bonner sphere~\cite{Verbus2016}.
Assuming an isotropic source\footnote{Actually, the \dd{} neutron flux varies by approximately a factor of two as a function of angle~\cite{Csikai1987}, but the isotropic assumption provides a convenient normalization.}, this corresponds to $(2.5 \pm 0.3) \times 10^{6}$~n/s into 4$\pi$ solid angle.
A total of 107.2~live-hours of \dd{} neutron data was acquired and used for the analysis.

\subsection{Beam energy purity cuts}

Monte Carlo simulation studies using LUXSim/GEANT4~\cite{AkeribBaiBedikianEtAl2012, AgostinelliAllisonAmakoEtAl2003} indicate that after selecting events using a cylindrical analysis volume in line with the neutron beam in the TPC, 95\% of accepted events are due to neutrons with energies within 6\% of the initial energy at the \dd{} source~\cite{Malling2014}.
This position cut requires that the first scatter has a reconstructed location of $y^{\prime} > 15$~cm and is within the 4.9~cm diameter of the neutron beam projection in the detector active region.
These position-based analysis cuts are referred to as the ``neutron energy purity cuts'' in the following sections.
Any residual electron recoil contamination produced by neutron capture or inelastic scatters in passive materials was identified and removed in the data analysis~\cite{Malling2014, Verbus2016}.
There are several xenon metastable states resulting from inelastic neutron scatters that do not produce a prompt electron recoil signal.
Contamination due to events arising from this type of inelastic process was calculated to be $<$1\% of the elastic nuclear recoil rate.
The systematic uncertainty in the reconstructed energy due to the variation in the atomic mass and cross-section over xenon isotopes with significant natural abundance was estimated to be $<$2\% for all energies---subdominant to other uncertainties in the following analyses.

\section{Low-energy ionization yield} \label{sec:dd_low_energy_qy}

The ionization yield was measured as a function of nuclear recoil energy from 0.7 to 24.2~\kevnr{} using neutrons that scatter twice in the active liquid xenon volume.

\subsection{Absolute measurement of nuclear recoil energy using double-scatter events} \label{sec:abs_energy_from_double_scatter}

For double-scatter neutron events, the scattering angle between the first and second interaction sites was calculated based upon the reconstructed ($x$,\,$y$,\,$z$) position of each site. 
The scattering angle in the center-of-mass frame, $\theta_{\textrm{CM}}$, is related to the recoil energy associated with the first interaction:

\begin{equation} \label{eq:recoil_energy_equation}
    E_{\textrm{nr}} = E_{n} \frac{4m_{n}m_{\textrm{Xe}}}{\left(m_{n} + m_{\textrm{Xe}}\right)^{2}} \frac{1-\cos{(\theta_{\textrm{CM}})}}{2} \,\text{,}
\end{equation}

\noindent
where $m_{\textrm{Xe}}$ is the average atomic mass of Xe, $m_{n}$ is the mass of the neutron, and $E_{n}$ is the energy of the incident neutron.
The relationship between $\theta_{\textrm{CM}}$ and the scattering angle in the laboratory frame, $\theta_{\textrm{lab}}$, is given by

\begin{equation} \label{eq:recoil_angle_cm_to_lab}
    \tan{\theta_{\textrm{lab}}} = \frac{\sin{\theta_{\textrm{CM}}}}{m_{n}/m_{\textrm{Xe}} + \cos{\theta_{\textrm{CM}}}} \,\text{.}
\end{equation}

\noindent
For the measurement presented here, the relationship 

\begin{equation} \label{eq:recoil_angle_approx_one}
    \frac{1-\cos{(\theta_{\textrm{lab}})}}{1-\cos{(\theta_{\textrm{CM}})}} \approx 1 
\end{equation}

\noindent
is accurate to better than 2\% for all scattering angles.

This absolute determination of the recoil energy combined with the observed $S2$ from the first interaction provides a direct \qy{} calibration. 
A conceptual schematic of this type of event is shown in Fig.~\ref{fig:lux_dd_scatter_diagram}.
The ($x$,\,$y$) positions were determined using the algorithm described in Ref.~\cite{SolovovBelovAkimovEtAl2012}.
The $z$ positions were measured using the ionization electron drift time.
The variable $\theta_{\textrm{lab}}$ was reconstructed using the measured 3D positions of the first and second interaction sites.
The ionization yield measurement used individual events with a reconstructed nuclear recoil energy between 0.3 and 30~\kevnr{}, which corresponds to a measured neutron scattering angle range of 7$^{\circ}$ to 79$^{\circ}$.
For comparison, the recoil energy spectrum endpoint produced by 180$^{\circ}$ neutron scatters corresponds to a nuclear recoil energy of 74~\kevnr{}.

\subsection{Recoil energy measurement uncertainties} \label{sec:sec:dd_low_energy_qy_recoil_energy_uncertainties}

The statistical uncertainty associated with the ($x$,\,$y$) position reconstruction is dependent upon the size of $S2$.
The typical statistical error in the reconstructed $x$ and $y$ coordinates was typically no more than $\sim$1~cm for the signal sizes used for this analysis, with a maximum statistical error of $\sim$2~cm at the 36~phd raw $S2$ threshold.
The systematic uncertainties in the reconstructed $x$ and $y$ positions were estimated to be 0.0--0.7~cm, with the best estimate of 0.35~cm~\cite{Akeribothers2016}.
The $z$ position of each scatter site has a statistical uncertainty of $\sim$0.1~cm~\cite{Faham2014}.
After neutron energy purity cuts, the incident direction of neutrons producing accepted events was parallel within 1$^{\circ}$ of the beam direction based upon the solid angle presented by the collimation conduit.
An estimated position uncertainty on the beam entry position into the TPC of 0.6~cm in $x^{\prime}$ and $z^{\prime}$ was included in the per-event energy determination.

The error in the measured nuclear recoil energy at the first scattering site in an individual double-scatter event was estimated by propagating the error on the $x$, $y$, and $z$ coordinates through to the reconstructed angle.
Events with larger distances between interaction sites and/or large separation between interactions along the $z$ direction have a smaller fractional error in the reconstructed event recoil energy. 
The per-event uncertainties on the reconstructed recoil energy were used to weight the events to optimize the fractional error on the mean reconstructed energy of a particular recoil energy bin.
The weighting scheme is described in detail in Ref.~\cite{Verbus2016}.

A detailed study was made of the way in which event reconstruction populates the measured nuclear recoil energy bins. 
Events with true energy outside a given bin can bleed inside, due to the non-zero resolution of the angle-based measurement.
This is an example of Eddington bias~\cite{Eddington1913, Eddington1940} and must be accounted for in the analysis.\footnote{Eddington bias is commonly confused with the more widely known Malmquist bias, which is a related effect~\cite{Teerikorpi2004}.}
This effect broadens the width of the measured charge distribution in a given bin.
If additionally the underlying spectrum is falling (rising), there is more bleeding into the bin from lower (higher) energies, causing a negative (positive) bias in the mean measured charge per unit recoil energy with respect to the true yield.
Due to the S2 threshold, there are more high-energy events that can be reconstructed down into a given low-energy bin than there are lower-energy events that can be reconstructed up into the same bin.
A Monte Carlo simulation of multiple scatter neutron events in the LUX detector was used to quantify and generate corrections for these effects due to position reconstruction uncertainty and to verify the angular reconstruction algorithms used for the data analysis.
The Monte Carlo also includes $S2$ resolution effects due to fluctuations associated with signal creation and recombination as modeled by NEST~v1.0.
The electron lifetime and extraction efficiency effects are binomially applied and also contribute to the simulated $S2$ resolution. 
The simulation is described in detail in Ref.~\cite{Verbus2016} and the associated systematic uncertainties after this correction is applied are reflected in the results reported in Table~\ref{tab:dd_qy_result}.
It is important to note that the Eddington bias correction was only applied to the mean recoil energy of the event population in each bin.
As a consequence, for the results reported in this section, the defined recoil energy bin boundaries and per-event reconstructed recoil energies are reported before any Eddington bias correction.

\subsection{Double-scatter event selection} \label{sec:dd_low_energy_qy_event_selection}

The double scatter event structure was described in Sec.~\ref{sec:abs_energy_from_double_scatter}.
Scintillation from both interaction sites was observed as a single combined S1 signal because the maximum time-of-flight of a 2.45~MeV neutron between scattering locations in the LUX active region is $\sim$30~ns, which is similar to the time constant associated with the S1 pulse shape in liquid xenon.
Similar to normal single-scatter TPC operation, the S1 pulse was used to provide a start-time $t_{0}$ in the double-scatter analysis allowing the reconstruction of the $z$ position of both scatters with respect to the liquid surface.

The analysis threshold for S2 identification is raw $S2 > 36$~phd (1.5~extracted electrons) prior to position-dependent corrections.
This is a lower threshold than was used for the WIMP search analysis~\cite{AkeribAraujoBaiEtAl2015}, which is possible due to the small number of accidental coincidence events that can pass as legitimate double-scatters.
More discussion on the accidental coincidence double-scatter events is provided in Sec.~\ref{sec:dd_low_energy_qy_data_analysis}.

Multiple neutron interactions at similar $z$ can be misidentified as single interactions if there is significant overlap in the S2 waveforms.
The intrinsic S2 pulse width for a single neutron interaction site is due to the length of the detector's luminescent gas gap.
There is an additional $z$ dependent contribution to the intrinsic S2 signal width due to the longitudinal diffusion of electrons drifting in the liquid xenon~\cite{Sorensen2011a}.
A cut on the root-mean-square of the charge arrival time (RMS width) within S2 pulses was used to preferentially reject overlapping S2 signals. 
The optimum value of this upper limit on the RMS width was determined to be 775~ns via simulation.
This cut accepts 99\% of true single-site interactions, while rejecting 69\% of combined multiple-site interactions.
The remaining events containing S2 pulses composed of combined multiple interaction sites contribute to the background of events described in Sec.~\ref{sec:dd_low_energy_qy_data_analysis}.

The reconstructed ($x^{\prime}$,\,$y^{\prime}$) position of the first scatter satisfied the neutron energy purity cuts discussed in Sec.~\ref{sec:lux_dd_experimental_setup}.
Forward scatters were defined as events where the identified second scatter had a $y^{\prime}$ position deeper into the liquid xenon along the beam path than the identified first scatter.
The Euclidean distance $\rho$ was defined as the separation of scatter sites in physical 3D space.
A cut ensuring $\rho>5$~cm removed events with dominant systematic bias in angle reconstruction due to position reconstruction uncertainties.

Maximum signal size cuts on $S1$ and $S2$ were used to reject electron recoil events.
The thresholds for these cuts were conservatively informed using NEST~v0.98 and NEST~v1.0 for electron recoil and nuclear recoil signal yields, respectively~\cite{SzydagisFyhrieThorngrenEtAl2013, LenardoKazkazManalaysayEtAl2015}.
The cut on the coincident S1 signal of $S1 < 300$~phd accepts $>$99\% of \dd{} neutron double-scatter events.
The cut $S2 < 5000$~phd, applied to both scatters in each event, accepts $>$99\% of all \dd{} neutron S2 pulses while rejecting all 39.6~\kevee{} gamma rays from inelastic neutron scatters on $^{129}$Xe.
The next lowest-energy gamma ray resulting from an inelastic scatter is due to the 80.2~\kevee{} excitation of $^{131}$Xe, which is well outside of the parameter space of interest.

A cut on $S2[y_{2}^{\prime}]$ was used to ensure a high efficiency for the detection of the combined S1 signal. 
A requirement was imposed that $S2[y_{2}^{\prime}]$~$>$~225~phd.
This minimum cut on $S2[y_{2}^{\prime}]$ ensured a 90\% efficiency for detecting the combined S1 for double-scatter events due to the S1 contribution from the second scatter alone.
This cut accepts $>70\%$ of underlying double-scatter nuclear recoils before other cuts are applied and has a constant efficiency as a function of the energy deposited by the first neutron scatter.

For double-scatter events with both interaction sites within the projection of the neutron conduit, there can be ambiguity as to which interaction occurred first.
A cut on $S2[y_{2}^\prime] < 1500$~phd was effective in removing events in which a first scatter with $\theta \sim 180$~degrees is followed by a second scatter in the cylinder of the beam at smaller $y^{\prime}$.
Monte Carlo studies demonstrated that this cut accepts 89\% of good candidate \dd{} neutron forward-scatter events while rejecting 95\% of potential events where the interactions may have been incorrectly ordered by the analysis.

The JENDL-4 nuclear database was used to calculate the efficiencies presented in this section~\cite{ShibataIwamotoNakagawaEtAl2011}.

\begin{figure}[!htbp]
    \begin{center}
        \includegraphics[width=0.48\textwidth]{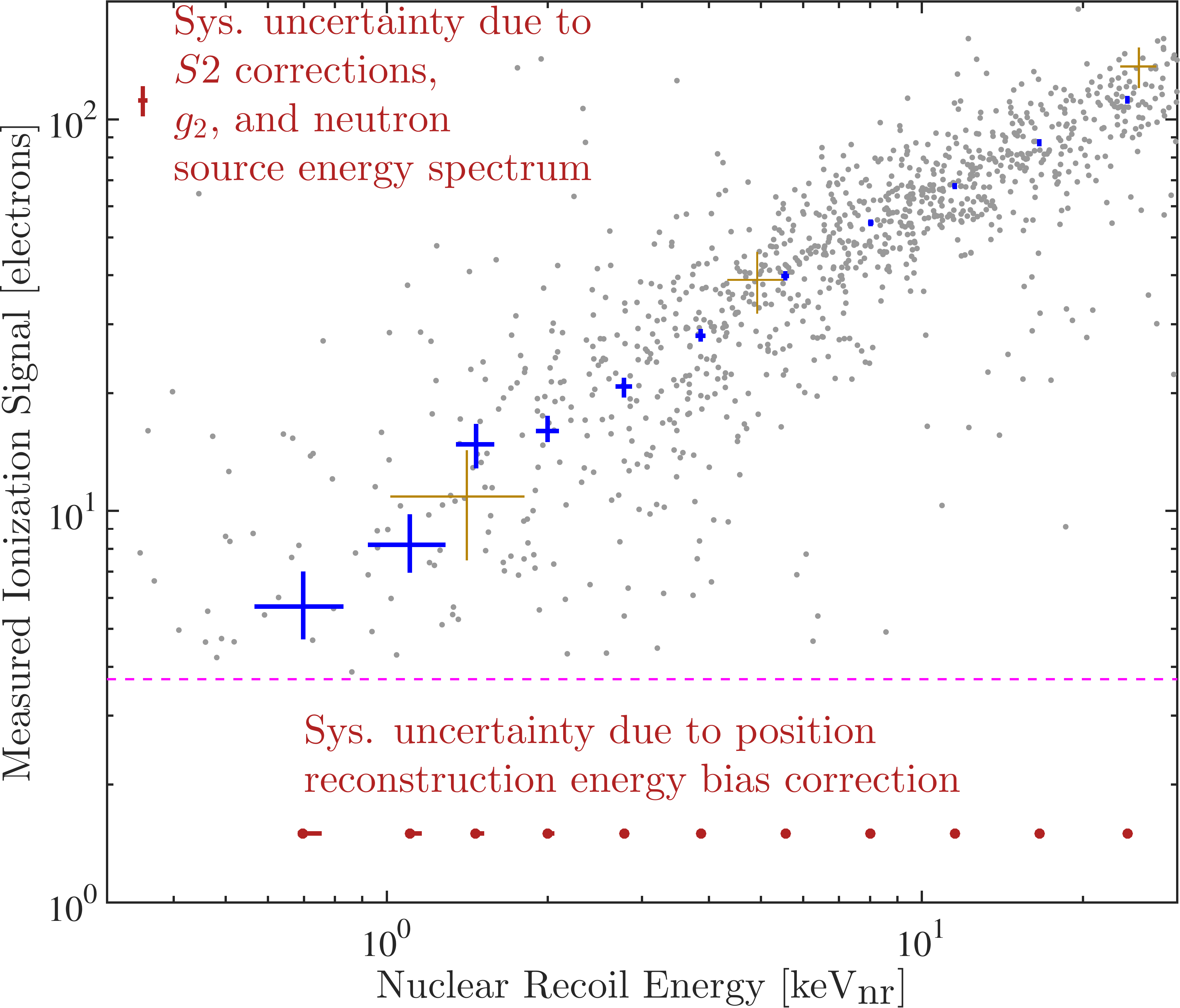}
        \vskip -0.1cm
        \caption{
            The gray points represent the measured ionization signal for each of the 1031 events remaining after all cuts in the double-scatter dataset.
            The gold crosses illustrate the estimated error associated with the most precisely measured individual events, both in ionization signal ($y$ error) and reconstructed energy ($x$ error).
            The measured ionization signal for each bin is represented by the blue crosses.
            As discussed in Sec.~\ref{sec:sec:dd_low_energy_qy_recoil_energy_uncertainties}, the mean recoil energy of the event population in each bin, represented by the location of the blue crosses on the horizontal axis, has been corrected for Eddington bias.
            The red error bars at the bottom of the plot represent the systematic uncertainty associated with this Eddington bias correction.
        }
        \vskip -0.5cm
        \label{fig:ionization_yield_ionization_signal_measured}
    \end{center}
\end{figure}

\subsection{Data analysis} \label{sec:dd_low_energy_qy_data_analysis}

\begin{figure}[!htbp]
    \begin{center}
        \includegraphics[width=0.46\textwidth]{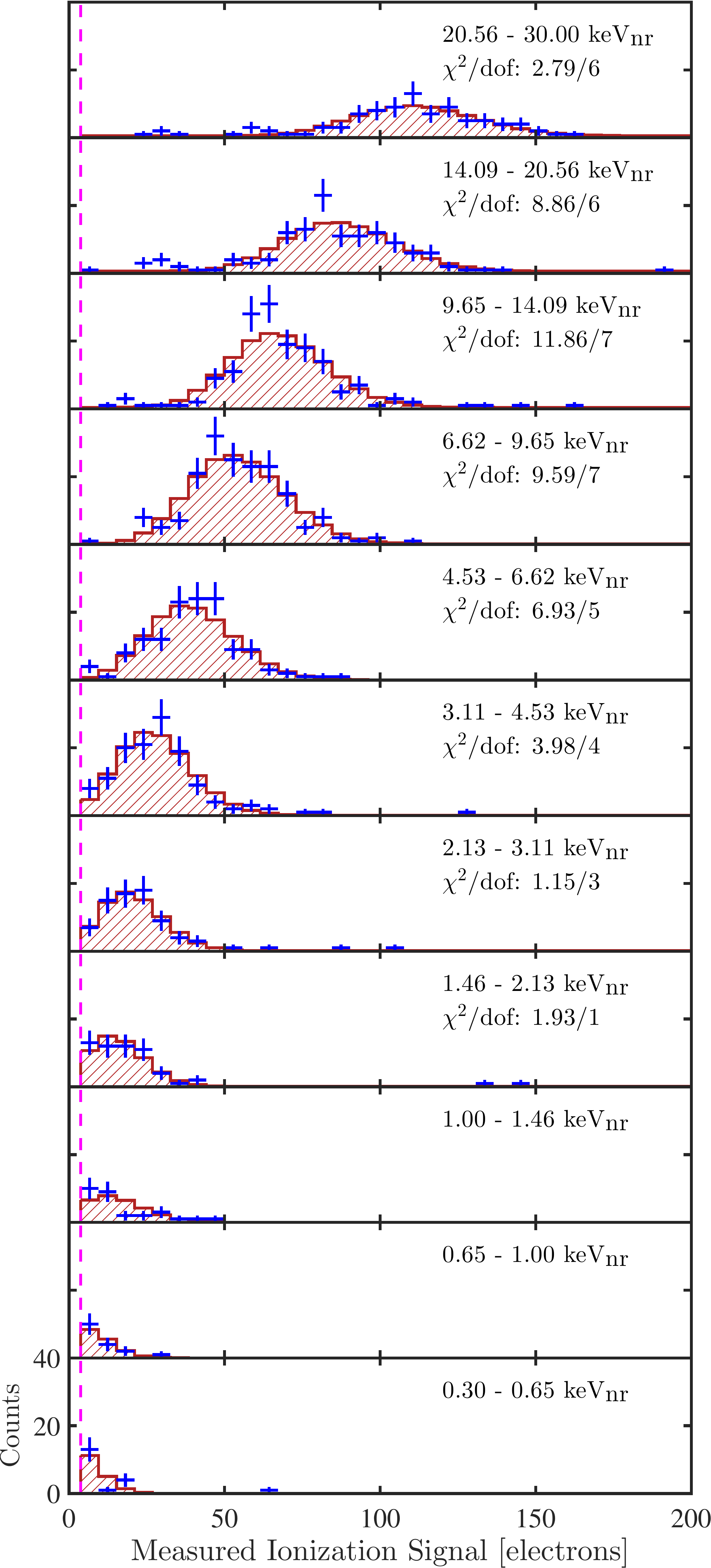}
        \vskip -0.1cm
        \caption{
            Histogram of the measured ionization signal with the best-fit model for each nuclear recoil energy bin.
            As discussed in Sec.~\ref{sec:sec:dd_low_energy_qy_recoil_energy_uncertainties}, the bin boundaries are defined based upon the per-event reconstructed nuclear recoil energy before the Eddington bias correction (the mean energy of events observed in each bin after the Eddington bias correction can be seen in Table~\ref{tab:dd_qy_result}).
            Data is shown by the blue crosses with Poisson error bars.
            The red shaded histogram is the best model fit to the data in each bin.
            The best-fit parameters were determined using an unbinned optimization.
            The ionization signal bins shown here were used to calculate $\chi^{2}$/dof values for energy bins where dof~$>$~0.
            The magenta line represents the approximate location of the S2 threshold.
            The axes limits are the same for each graph.
        }
        \vskip -0.5cm
        \label{fig:ionization_yield_ionization_signal_bins_all}
    \end{center}
\end{figure}

The per-event ionization signal is defined as the number of electrons \nel{} escaping recombination with ions at the interaction site. 
The ionization signal was determined for each event by dividing the position-corrected $S2[y_{1}^\prime]$ by the electron extraction efficiency and by the measured single electron size.
The uncertainty on the single electron size is subdominant ($\ll$1\%) to other uncertainties in the \qy{} analysis.
The 1031 events remaining after the application of all cuts are shown as gray points in Fig.~\ref{fig:ionization_yield_ionization_signal_measured}.
These events were divided into eleven \kevnr{} bins.
The two lowest-energy bins span the regions from 0.3--0.65~\kevnr{} and 0.65--1.0~\kevnr{}, respectively.
The remaining nine bins are logarithmically spaced from 1--30~\kevnr{}.
Histograms of the measured distribution of electrons escaping the interaction site for each bin are shown in Fig.~\ref{fig:ionization_yield_ionization_signal_bins_all}.

In order to determine the energy dependence of the charge yield, the analysis took full account of the statistical fluctuations associated with the ionization signal measurement, and the influence of the S2 threshold.
Given an input mean number of ionization electrons that escape recombination, a Monte Carlo based model was used to generate the expected probability distribution of the number of reconstructed electrons at the interaction site. 
The model is composed of an underlying Poisson distribution convolved with a Gaussian to account for the observed resolution of the ionization distribution.
Detector-specific effects including SE size and S2 threshold are included in the model.
Liquid xenon purity and electron extraction efficiency effects were applied binomially to the modeled number of ionization electrons to determine the distribution of observed electrons in the xenon gas.

\begin{figure}[!htbp]
    \begin{center}
        \includegraphics[width=0.48\textwidth]{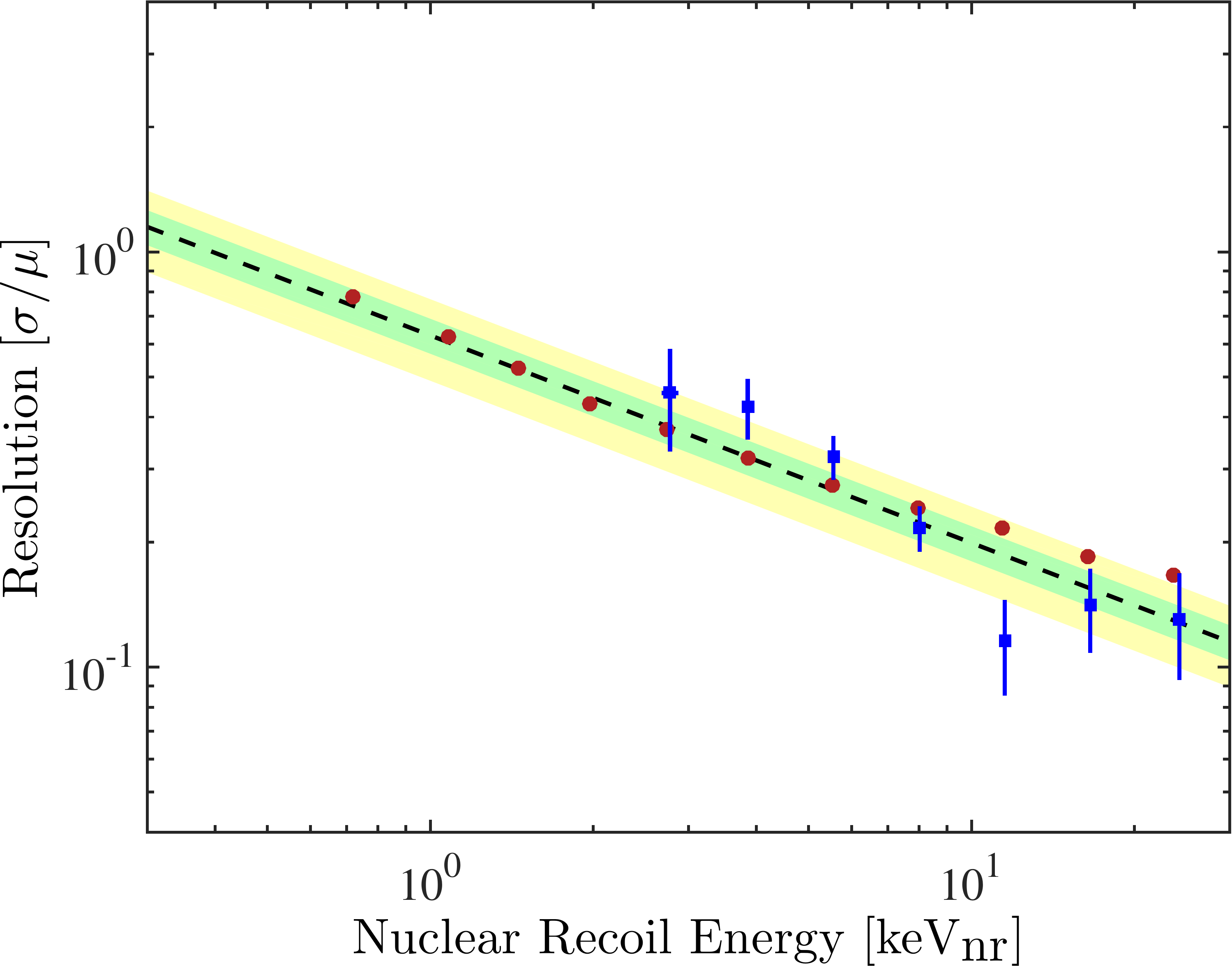}
        \vskip -0.1cm
        \caption{
            The measured resolution, $R$, of the ionization distributions in the seven highest-energy bins of the double-scatter dataset is represented by the blue squares.
            The estimated uncertainty in the resolution due to the extraction efficiency is a constant 4\% for all energies.
            The error bars are symmeterized for the fit following the procedure in Ref.~\cite{DAgostini2004}.
            The simulated resolution of the ionization distribution produced by a NEST~v1.0 Monte Carlo with modeled position reconstruction uncertainties is represented by the red circles.
            The black dashed line represents the best-fit to the blue squares given by $R_{0} = 0.64 /\sqrt{E_{\textrm{nr}} / \mbox{\kevnr{}}}$.
            The fit has a $\chi^{2}$/dof = 10.6/6, which corresponds to a p-value of 0.12.
            The one and two sigma contours on the parameter $a$ are shown in green and yellow, respectively.
        }
        \vskip -0.5cm
        \label{fig:omega_data_measurement_ionization_distribution_resolution_no_sys}
    \end{center}
\end{figure}

The most significant contribution to the resolution of the ionization distribution is Eddington bias.
This arises from uncertainty in reconstructed energy due to the position reconstruction effects described in Sec.~\ref{sec:abs_energy_from_double_scatter}.
The expected ionization resolution after Eddington bias effects were addressed was confirmed to have an energy dependence ${\propto} 1/\sqrt{E_{\textrm{nr}}}$ via simulation~\cite{Verbus2016}.
The resolution in the model, set using the variance of the Gaussian convolution, was determined by fitting the signal model to the seven highest-energy \qy{} bins where S2 threshold effects are minimal as shown in Fig.~\ref{fig:ionization_yield_ionization_signal_bins_all}.
The $a/\sqrt{E_{\textrm{nr}}}$ dependence was fit to the measured ionization resolution for these seven bins as shown in Fig.~\ref{fig:omega_data_measurement_ionization_distribution_resolution_no_sys}.
The value of the parameter $a \pm \sigma_{a}$ was measured to be $0.64 \pm 0.06~\sqrt{\mbox{\kevnr{}}}$.
The mean of the signal model distribution was an unconstrained nuisance parameter during this maximum-likelihood fit to extract the resolution.
The resulting additional uncertainty from this nuisance parameter is reflected in the reported error bars.

After determining the nuclear recoil energy dependence of the ionization signal resolution, the final signal model was fit to each bin.
The resulting ionization signal distribution and best-fit model for each bin is shown in Fig.~\ref{fig:ionization_yield_ionization_signal_bins_all}.
The ionization signal model was fit to the observed ionization distribution for each bin using an extended unbinned maximum likelihood optimization, with the modeled resolution implemented as a constrained nuisance parameter~\cite{Barlow1990}.
The log-likelihood for the optimization is

\begin{multline} 
    \ln{L} = \\
    -(N_{s} + N_{b}) - \ln{(N!)} + \ln{\left[\frac{1}{\sqrt{2\pi}\sigma_{R}}e^{- \frac{(R-R_{0})^{2}}{2\sigma_{R}^{2}}}\right]} + \\
    \sum\limits_{i=1}^{N}\ln{\left[ N_{s}p_{s}(x_i \vert n_{e}, R) + N_{b}p_{b}(x_i) \right]} \,\text{,}
    \label{eq:likelihood_qy_fit}
\end{multline}

\noindent
where the parameters $N_{s}$, $N_{b}$, $n_{e}$, and $R$ are varied in the optimization.
The index $i$ iterates over each event $x_{i}$ in the particular \kevnr{} bin, and $N$ is the total observed number of events in the bin.
The parameter $N_{s}$ is the number of signal events, and $N_{b}$ is the number of background events in the fit.
The parameter of primary interest is $n_{e}$, the measured number of ionization electrons escaping recombination with ions at the interaction site.
The parameter $R$ is the resolution of the reconstructed electron distribution at the interaction site.
The parameters $n_{e}$ and $R$ are inputs to the signal model PDF $p_{s}(x_i \vert n_{e}, R)$, where $R$ functions as a nuisance parameter constrained by the measured resolution best-fit to the seven highest-energy bins shown in Fig.~\ref{fig:omega_data_measurement_ionization_distribution_resolution_no_sys}.
This constraint on $R$ is enforced using the parameters $R_{0}$ and $\sigma_{R}$ in Eq.~\ref{eq:likelihood_qy_fit} for each reconstructed energy bin.
For each recoil energy bin, these resolution parameters are

\begin{equation}
    R_{0} = a/\sqrt{E_{\textrm{nr}}}
\end{equation}

\noindent
and

\begin{equation}
    \sigma_{R} = \sigma_{a}/\sqrt{E_{\textrm{nr}}} \,\text{.}
\end{equation}

\noindent
The parameters $a$ and $\sigma_{a}$ were defined earlier based upon the fit in Fig.~\ref{fig:omega_data_measurement_ionization_distribution_resolution_no_sys}.

Events outside the main peak were accommodated by a flat continuum background PDF $p_{b}(x_i)$.
The classes of events contributing to this background are discussed in Sec.~\ref{sec:dd_low_energy_qy_background_and_uncertainties}.

\begin{figure}[!htbp]
    \begin{center}
        \includegraphics[width=0.48\textwidth]{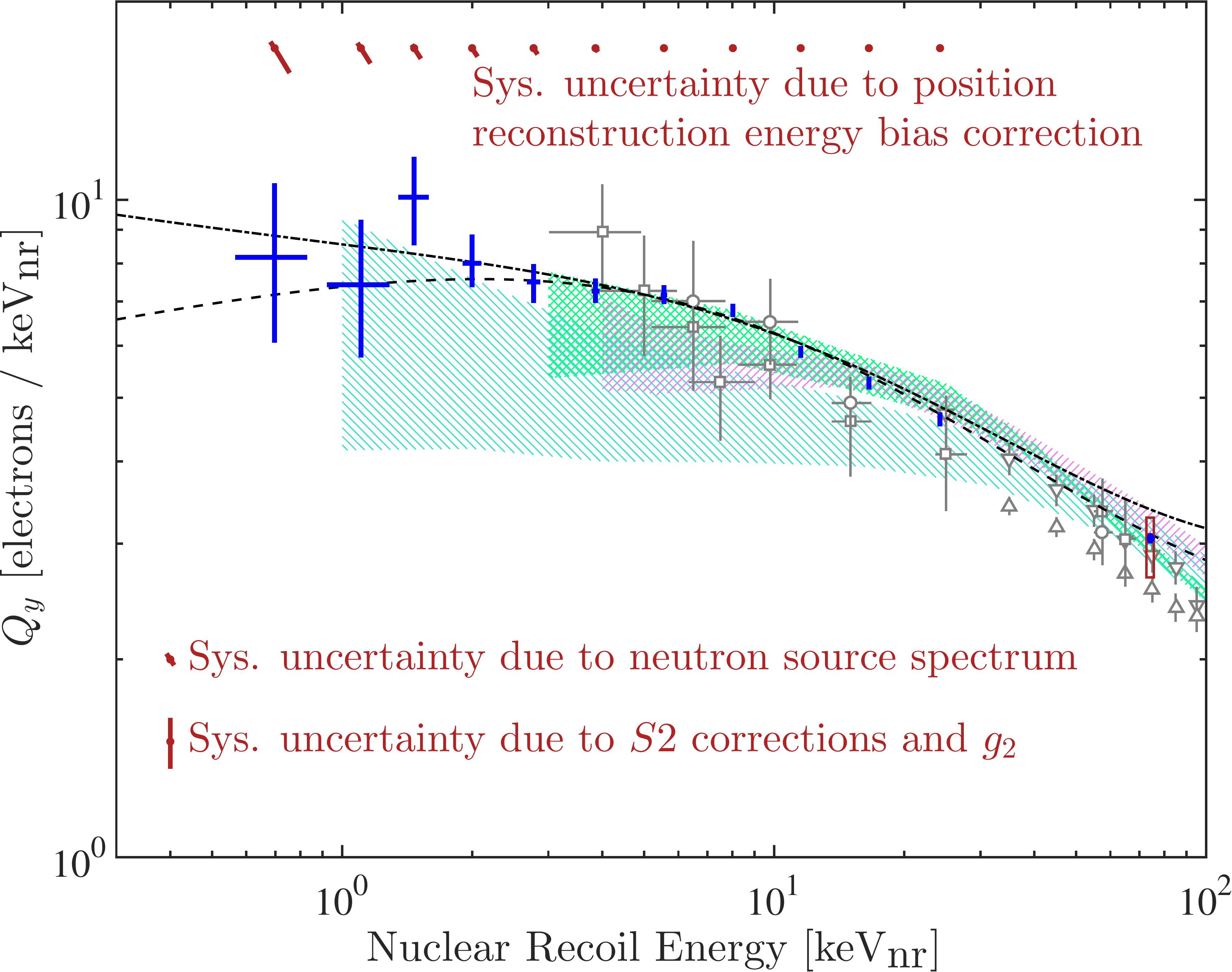}
        \vskip -0.1cm
        \caption{
            The LUX measured low-energy ionization yield at 180 V/cm is represented by the blue crosses.
            The red error bars at the bottom left of the plot represent systematic uncertainties with a constant scaling across all points, including the uncertainty in the mean neutron energy from the \dd{} source, $S2$ position-based corrections, and the LUX measured $g_2$.
            The red error bars at the top of the plot represent the systematic uncertainty associated with the Eddington bias correction for the mean energy of each bin. 
            The red box represents the associated systematic uncertainty on the measured endpoint yield at 74~\kevnr{}.
            The gray data points represent other angle-based measurements with an absolute energy scale.
            The gray squares ($\Box$) and circles ($\bigcirc$) correspond to measurements at 1 kV/cm and 4 kV/cm, respectively~\cite{Manzur2010}.
            The gray triangles were measured at 0.3~kV/cm ($\triangledown$) and 0.1~kV/cm ($\triangle$)~\cite{Aprile2006a}.
            The hatched bands represent simulated-spectrum-based measurements with a best-fit energy scale.
            The purple single right-hatched ($///$) band was measured at an average field of 3.6 kV/cm~\cite{Horn2011}.
            The teal single left-hatched ($\backslash\backslash\backslash$) band corresponds to a measurement at 730~V/cm~\cite{Sorensen2010a}.
            The green cross-hatched band was measured at 530~V/cm~\cite{AprileAlfonsiArisakaEtAl2013}.
            The dashed (dot-dashed) black line corresponds to the Lindhard-based (Bezrukov-based) LUX best-fit NEST model described in Sec.~\ref{sec:nest_post_dd}.
        }
        \vskip -0.5cm
        \label{fig:ionization_yield_qy_endpoint_ionization_yield}
    \end{center}
\end{figure}

The ionization signal model best-fit for each of the eleven bins is shown in Fig.~\ref{fig:ionization_yield_ionization_signal_measured}.
The corresponding measured ionization yield is shown in Fig.~\ref{fig:ionization_yield_qy_endpoint_ionization_yield}.
The ionization yield was calculated from the mean ionization signal shown in Fig.~\ref{fig:ionization_yield_ionization_signal_measured} by dividing each point by the reconstructed nuclear recoil energy to obtain electrons per \kevnr{}.
The measured ionization yield and associated per-bin uncertainties are shown in Table~\ref{tab:dd_qy_result}.

\begin{table}[!htbp]
    \centering
    \caption{
        Measured ionization yield for nuclear recoils in liquid xenon at 180~V/cm and associated $1\sigma$ statistical uncertainties.
        The systematic uncertainty in energy due to the correction for Eddington bias is denoted by $\Delta E_{\textrm{nr}}/ E_{\textrm{nr}}$.
        This uncertainty in energy is represented in Fig.~\ref{fig:ionization_yield_qy_endpoint_ionization_yield} by a slanted error bar. 
    }
    \label{tab:dd_qy_result}
    \setlength{\extrarowheight}{.5em}
    \begin{tabular*}{\columnwidth}{S @{\extracolsep{\fill}} Sl}
        \hline \hline
        {$E_{\textrm{nr}}$} & {\qy{}} & {$\Delta E_{\textrm{nr}}/ E_{\textrm{nr}}$} \\
        {(\kevnr{})} & {(e$^{-}$/\kevnr{})} & {(\%)} \\ 
        \hline
        0.70~$\pm$~0.13 & 8.2 $^{+2.4}_{-2.1}$ & $^{+8}_{-2}$ \\
        1.10~$\pm$~0.18 & 7.4 $^{+1.9}_{-1.7}$ & $^{+5}_{-1.9}$ \\
        1.47~$\pm$~0.12 & 10.1 $^{+1.5}_{-1.6}$ & $^{+3}_{-1.3}$ \\
        2.00~$\pm$~0.10 & 8.0 $^{+0.9}_{-0.6}$ & $^{+2}_{-1.3}$ \\
        2.77~$\pm$~0.10 & 7.5 $^{+0.5}_{-0.5}$ & $^{+2}_{-0.7}$ \\
        3.86~$\pm$~0.08 & 7.3 $^{+0.3}_{-0.3}$ & $^{+1.3}_{-0.5}$ \\
        5.55~$\pm$~0.09 & 7.2 $^{+0.2}_{-0.2}$ & $^{+0.7}_{-0.2}$ \\
        8.02~$\pm$~0.10 & 6.8 $^{+0.15}_{-0.17}$ & $^{+0.16}_{-0.05}$ \\
        11.52~$\pm$~0.12 & 5.88 $^{+0.12}_{-0.13}$ & $^{+0.13}_{-0.3}$ \\
        16.56~$\pm$~0.16 & 5.28 $^{+0.11}_{-0.13}$ & $^{+0.2}_{-0.7}$ \\
        24.2~$\pm$~0.2 & 4.62 $^{+0.13}_{-0.10}$ & $^{+0.4}_{-1.0}$ \\
        \hline \hline
    \end{tabular*}
\end{table}

To verify the consistency of the measured yields with the observed absolute event rate, we performed a LUXSim/GEANT4 based neutron double-scatter simulation using the NEST model described in Sec.~\ref{sec:nest_post_dd}.
This simulation used a model of the full calibration conduit geometry with the neutron source external to the water tank. 
Simulated per-channel waveforms were produced for each Monte Carlo event.
The simulated waveform data were reduced using the standard experimental LUX \dd{} data processing and analysis pipeline.

\begin{figure}[!htbp]
    \begin{center}
        \includegraphics[angle=90,width=0.48\textwidth]{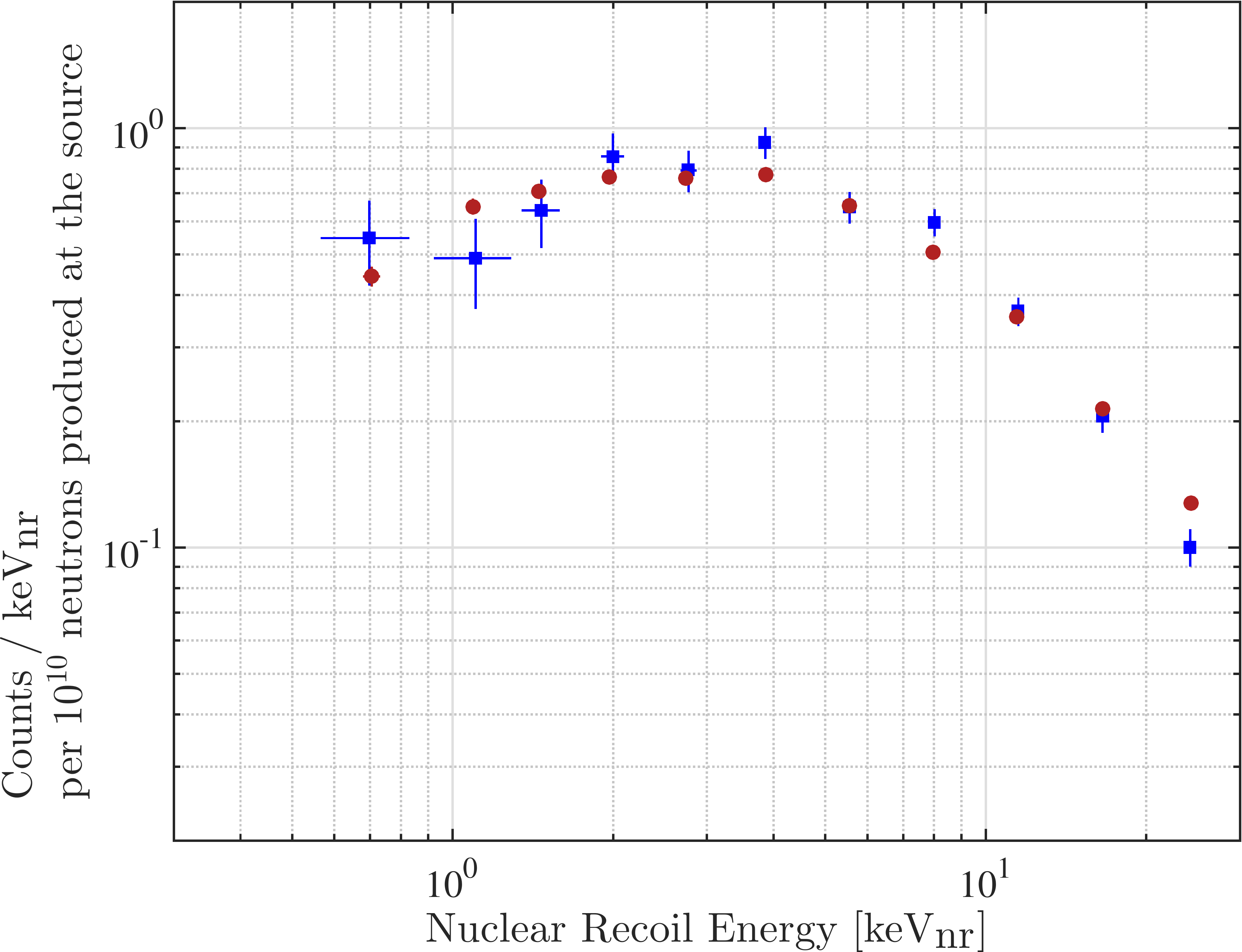}
        \vskip -0.1cm
        \caption{
            The observed rate of double-scatter neutron events in the \qy{} analysis is represented by the blue squares.
            An identical analysis of simulated waveforms produced by LUXSim/GEANT4 using the LUX measured nuclear recoil signal yields was performed.
            The results are shown as red circles.
            The simulation statistical error bars are smaller than the size of the data points unless otherwise depicted.
            The results are normalized by the number of neutrons produced by the \dd{} source outside the water tank.
            The $\chi^{2}$/dof value is 14.6/10 based upon statistical uncertainties only, which yields a p-value of 0.15.
        }
        \vskip -0.5cm
        \label{fig:ionization_yield_full_luxsim_double_scatter_num_events_data_and_sim}
    \end{center}
\end{figure}

The event rate in each \qy{} analysis bin is shown in Fig.~\ref{fig:ionization_yield_full_luxsim_double_scatter_num_events_data_and_sim} for both data and simulation.
The data and simulation results were normalized by the total number of neutrons produced at the \dd{} source outside the water shield.
For consistency with the other yield results, the simulation data points were updated to use the more modern angular scattering cross-sections from the JENDL-4 nuclear databases instead of G4NDL3.14.
The absolute value of the correction factor was $\leq$1\% for energy bins up to 5.55~\kevnr{} and was a maximum of 5\% at 24.2~\kevnr{}. 
The best agreement was achieved assuming an isotropic neutron source rate of $2.6 \times 10^{6}$~n/s for the data normalization, which is in agreement with the independently measured source rate of $(2.5 \pm 0.3) \times 10^{6}$~n/s.
This agreement between the data and simulation in both absolute rate and shape confirms the consistency of the LUX \dd{} measured yields and the number of events seen in the double-scatter data at nuclear recoil energies as low as 0.7~\kevnr{}.

\subsection{Background and uncertainties} \label{sec:dd_low_energy_qy_background_and_uncertainties}

There are six classes of events that contribute to the continuum background observed outside of the signal peaks in Fig.~\ref{fig:ionization_yield_ionization_signal_bins_all}.
The best-fit number of background events, $N_{b}$, accounts for less than 6\% of the area in the first nine recoil energy bins and less than 20\% in the three highest-energy bins.
The common quality of continuum background events is that the measured angle is not directly related to the true recoil energy at the first scattering site.
\begin{enumerate}[i.]
    \item The first class consists of three or more scatter events classified as two scatter events, because the pulse finding algorithm combines two S2 pulses that are close in $z$ position.
        The S2 pulse width cut preferentially removes events with combined S2 signals, while having an average acceptance of 94\% for legitimate double-scatters after all other cuts are applied.
        The corresponding acceptance of legitimate double-scatter events with a first interaction nuclear recoil energy of $<$2~\kevnr{} and $<$1~\kevnr{} is 86\% and 80\%, respectively. 
    \item The second class contains events that have $>$2~scatters, but only two of the scatters are above the 36~phd raw S2 threshold.
        As was also the case for the first class of events, if this was a dominant effect the observed mean path length between scatters would be longer than expected based upon the mean free path of 2.45~MeV neutrons in liquid xenon.
        The distribution of the measured distance between interactions in double-scatter events was demonstrated to be consistent with simulation using neutron scattering cross-sections from JENDL-4~\cite{Verbus2016}.
    \item The third class consists of events that scatter once within the neutron beam projection in the TPC, then scatter in passive detector materials, and then finally scatter again in the active liquid xenon volume.
        This is effectively a 3+ scatter event that is identified as a two scatter event.
    \item The fourth class of events is the accidental coincidence of delayed electron emission (SE or small S2) classified as the first scatter, with a legitimate single-scatter neutron event classified as the second scatter.
        (The reverse process is suppressed due to the minimum pulse area requirement for the second scatter, which corresponds to $\sim$9 extracted electrons.)
        The measured background rate of random small S2 pulses indicates that $<$0.1\% of events in the \qy{} dataset after the analysis volume constraints and pulse area thresholds for $S2[y_{1}^{\prime}]$ and $S2[y_{2}^{\prime}]$ are this type of accidental coincidence.
    \item The fifth class of events are produced by the small number of incident neutrons that have lost a significant fraction of their energy in passive detector materials but pass the energy purity cuts.
        The nuclear recoil energy bins are determined by scattering angle, so this is a unidirectional effect that could produce a $\sim$5\% excess of events at lower ionization signal in a given bin.
        It is possible that some evidence of this effect is seen in the high-energy bins in Fig.~\ref{fig:ionization_yield_ionization_signal_bins_all}.
        It is worth noting that due to the angle-based energy scale neutrons that have lost energy in passive materials can only suppress the measured charge yield.
    \item The sixth class contains events where the order of the first and second neutron interactions was incorrectly defined.
        The double-scatter event selection criteria described in Sec.~\ref{sec:dd_low_energy_qy_event_selection} ensures that contamination of this type is negligible.
\end{enumerate}

Table~\ref{tab:dd_qy_result} contains the statistical errors for the reconstructed energy and the measured \qy{} as returned by the maximum-likelihood optimization.
The reported errors on the measured \qy{} values were extracted from the log-likelihood contour accounting for variations in all four parameters in the fit.
The third column contains the systematic uncertainty in energy due to the Eddington bias correction.
Systematic uncertainties common to all bins in the low-energy ionization yield measurement and endpoint \qy{} measurement are listed in Table~\ref{tab:dd_common_qy_uncertantites}.

The systematic uncertainty in \qy{} due to the S2 threshold was confirmed to be subdominant to other quoted uncertainties by varying the modeled threshold by 10\% and repeating the fitting procedure.
The low-energy \qy{} analysis was repeated using a smaller fiducial analysis volume ensuring that $r < 21$~cm and $30 < \text{drift time} < 290$~$\mu$s to test potential systematic effects associated with the choice of analysis volume.
The results of this check for systematic effects are within the quoted $1\sigma$ statistical uncertainties in Table~\ref{tab:dd_qy_result}. 

\begin{table}[!htbp]
    \centering
    \caption{
        Uncertainties common to the \qy{} measurement both at low energies and at the \dd{} recoil spectrum endpoint.
        The second column lists systematic uncertainties associated with the mean reconstructed ionization signal \nel{}.
        The third column lists systematic uncertainties associated with the mean reconstructed energy, $E_{\textrm{nr}}$.
        Quoted uncertainties are symmetric ($\pm$) unless otherwise indicated.
    } 
    \label{tab:dd_common_qy_uncertantites}
    \setlength{\extrarowheight}{.5em}
    \begin{tabular*}{\columnwidth}{l @{\extracolsep{\fill}} cc}
        \hline \hline
        Source of Uncertainty & $\Delta$\nel{}/\nel{} & $\Delta E_{\textrm{nr}} / E_{\textrm{nr}}$ \\
        & (\%) & (\%) \\
        \hline
        SE size & $\ll$1 & - \\
        e$^{-}$ extraction efficiency & 8 & - \\
        $S2$ correction (3D position) & 2.5 & - \\
        $S2$ correction (non-uniform field) & $^{+0}_{-2.5}$ & - \\
        Mean neutron energy from source & - & 2 \\
        \hline
        Total & $^{+8}_{-9}$& 2 \\
        \hline \hline
    \end{tabular*}
\end{table}

\section{Low-energy scintillation yield} \label{sec:dd_low_energy_ly}

The \qy{} result provides a precise \insitu{} measurement of the charge yield as a function of energy, which defines the $S2$ response as a function of recoil energy between 0.7--74~\kevnr{} (the endpoint yields at 74~\kevnr{} were obtained as described in Sec.~\ref{sec:dd_endpoint}).
The single-scatter (one S1 and one S2) event population was then used to calibrate the $S1$ yield using the observed $S2$ as a measure of energy.
A single-scatter signal model that simultaneously provides simulated $S1$ and $S2$ distributions was developed as described in Sec.~\ref{sec:dd_low_energy_ly_signal_model}.
The ionization yield in the model was fixed to the measured \qy{} from Sec.~\ref{sec:dd_low_energy_qy}, while \ly{} was varied and the output compared to data to extract the best-fit scintillation yield for 1.08--12.8~\kevnr{} nuclear recoils.
Because the measurements of the signal yields are performed \insitu{}, the uncertainty in $g_2$ does not contribute to the uncertainty in the \ly{} energy scale.
The precisely measured $g_1$ value is used to directly report the light yield in the absolute units of photons/\kevnr{}.

We use the model described in Sec.~\ref{sec:dd_low_energy_ly_signal_model} to measure the light yield for energies as low as 1.08~\kevnr{}, where only a fraction of the events are above the S1 and S2 detection thresholds.
The main challenges in this regime are ensuring that the thresholds and resolution are well modeled for both $S1$ and $S2$.
The LUX S1 and S2 threshold behavior is well understood~\cite{AkeribAraujoBaiEtAl2015, Akeribothers2016} and is included in the simulation used for best-fit parameter estimates.
Uncertainty in the measured \ly{} due to uncertainties in the modeled S1 and S2 thresholds are quantified and discussed in Sec.~\ref{sec:dd_low_energy_ly_background_and_uncertainties}.
The $S1$ resolution is dominated by Poisson fluctuations in the number of detected photons. 
The $S2$ resolution due to the Fano factor associated with quanta production, recombination fluctuations, and detector effects (purity, electron extraction) is constrained by the results in Sec.~\ref{sec:dd_low_energy_qy} and is consistent with the Poisson expectation of the model over the energy range spanned by the reported \ly{} data points.
The shape of the $S1$ vs. $S2$ distribution in data and simulation is compared in Fig.~\ref{fig:final_single_scatter_sim_ly_plots_final_single_scatters_s1_vs_s2_subplots}.

\subsection{Single-scatter event selection} \label{sec:dd_low_energy_ly_event_section}

The single-scatter pulse pairing requires an identified S1 pulse preceding an identified S2 pulse.
The S1 identification threshold requires a coincidence of 2~PMTs each with signal $>$0.25~phd.
The S2 analysis threshold required that raw $S2 > 55$~phd to reduce the number of potential accidental coincidence events.
This S2 threshold is higher than that used for the low-energy \qy{} analysis described in Sec.~\ref{sec:dd_low_energy_qy} to ensure rejection of accidental coincidences masquerading as single scatters.
The LUX WIMP search analysis used a higher raw $S2 > 164$~phd threshold due to the longer exposure and lower signal to accidental coincidence background event ratio.
The origin and measured residual number of accidental coincidence events are discussed later in Sec.~\ref{sec:dd_low_energy_ly_background_and_uncertainties}.
The same maximum area thresholds used in the low-energy \qy{} analysis are applied to $S1$ and $S2$.

The lower S2 threshold compared to that used for the WIMP search analyses provides an increased efficiency for the detection of single-scatter events associated with low-energy nuclear recoils.
The efficiencies for detecting 1~\kevnr{} and 2~\kevnr{} nuclear recoils were estimated after all analysis cuts to be 4\% and 25\%, respectively.
In addition to this increased detection efficiency due to the lower S2 threshold, the underlying true nuclear recoil spectrum produced by 2.45~MeV neutrons in liquid xenon sharply increases at low energies, which provides an additional enhancement in the relative number of low-energy events.

The neutron beam energy purity cuts were applied ensuring that only single-scatters within the 4.9~cm beam pipe projection with $y^{\prime} > 15$~cm were accepted.
A radial position cut ensuring $r < 21$~cm was applied.

Data quality cuts were applied to remove events due to accidental triggers in the period of delayed electron extraction, photoionization of impurities in the liquid xenon, or other photoelectric feedback effects following large S2 pulses that can span many subsequent event windows.
An upper limit on the total raw pulse area in the event record outside of the identified S1 and S2 of 219~phd was applied.
A cut ensuring that there are no SE or S2 pulses in the event record before the identified single-scatter S1 pulses was applied to ensure quiet detector conditions in the period preceding the identified single scatter.
These requirements are independent of the nuclear recoil energy of the event, and accept 83\% of events after all other cuts are applied.
The same upper limit on the width of S2 pulses used in the \qy{} analysis was enforced to reject multi-site events at similar $z$.

\begin{figure}[!htbp]
    \begin{center}
        \includegraphics[width=0.48\textwidth]{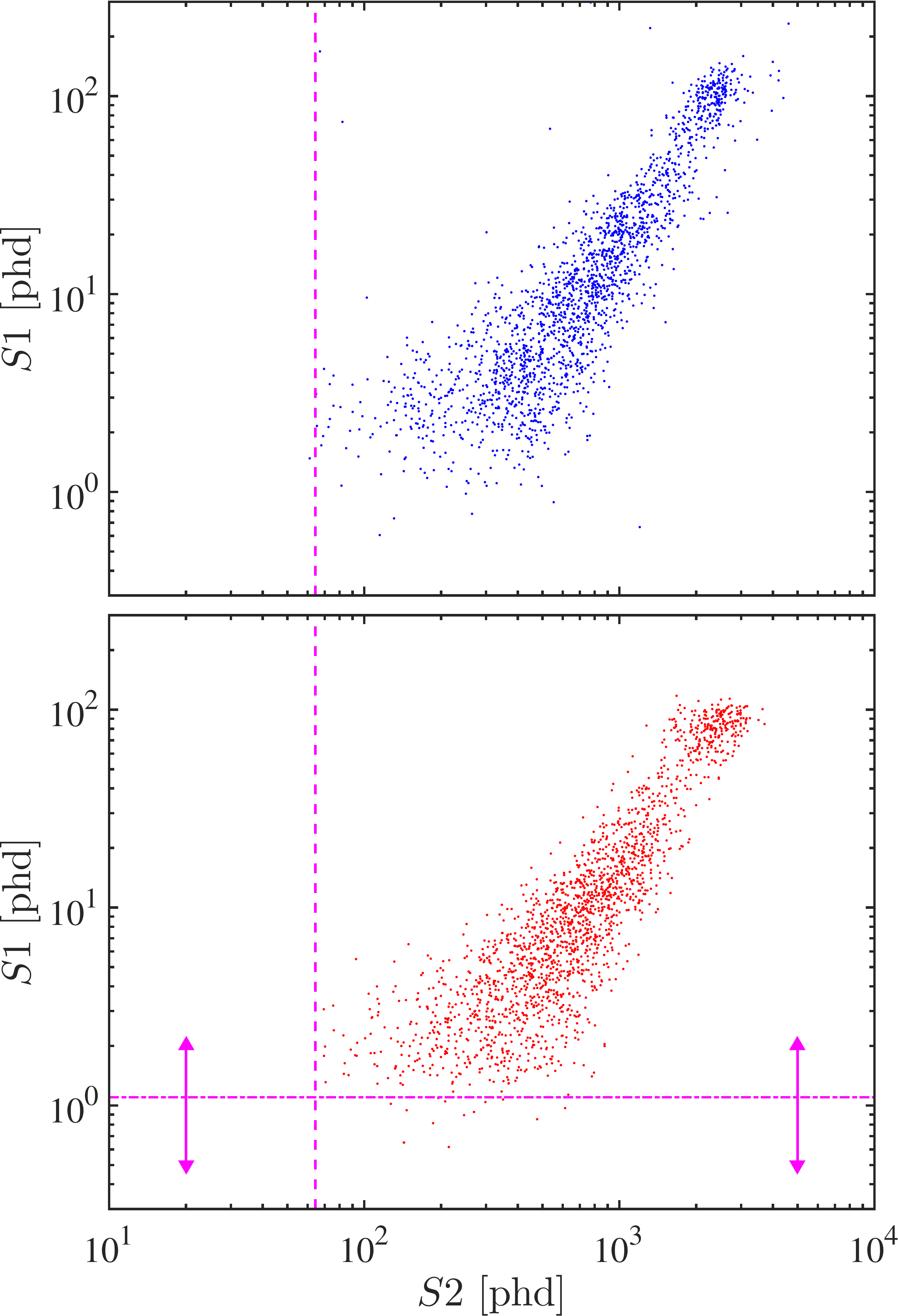}
        \vskip -0.1cm
        \caption{
            $S1$ vs. $S2$ single-scatter distribution for the \ly{} measurement.
            The 1931 events in data after all \ly{} analysis cuts are shown in this plot in blue in the upper frame.
            The non-uniformity of the distribution is due to the shape of the differential scattering cross-section~\cite{Verbus2016}.
            For comparison, a randomly selected sample consisting of the same number of simulated events, produced by the Lindhard-based NEST model described in Sec.~\ref{sec:nest_post_dd}, is shown in red in the lower frame.
            The raw $S2 > 55$~phd threshold is represented by the vertical dashed magenta line in corrected $S2$ space ($\sim$64~phd).
            The modeled S1 threshold requires that two photons contribute signal in the PMTs and that the resulting summed area is above the horizontal dot-dashed magenta line.
            This cutoff is varied over the range indicated by the magenta arrows, and the analysis is repeated to estimate the systematic uncertainty due to S1 threshold effects.
        }
        \vskip -0.5cm
        \label{fig:final_single_scatter_sim_ly_plots_final_single_scatters_s1_vs_s2_subplots}
    \end{center}
\end{figure}

After all event selection, position, and data quality cuts for the scintillation yield analysis were applied, a population of 1931 events remained in the neutron beam projection analysis volume.
The single-scatter event population is exceptionally clean with only a few events ($\ll$1\%) lying outside the main distribution as can be seen in Fig.~\ref{fig:final_single_scatter_sim_ly_plots_final_single_scatters_s1_vs_s2_subplots}.

\subsection{Signal model used for single-scatter simulation} \label{sec:dd_low_energy_ly_signal_model}

A Monte Carlo model of the S1 and S2 signal production was used to generate a simulated single-scatter event population to compare to the observed single-scatter events in data passing the cuts described in Sec.~\ref{sec:dd_low_energy_ly_event_section}.
The ionization yield in the model was fixed to match the LUX \dd{} measured \qy{}.
The modeled scintillation yield can be arbitrarily scaled using the free parameter $\xi$.
The model includes anti-correlation between $S1$ and $S2$ as well as fluctuations in exciton and ion creation, recombination, and biexcitonic quenching.
The variation of the scintillation yield in the model is achieved by scaling the number of photons produced at the interaction site before these fluctuation effects are applied. 
The JENDL-4.0 nuclear database was used to generate the underlying single-scatter nuclear recoil energy spectrum.

The S1 threshold in data required that two PMTs independently observe a signal greater than 0.25~phd.
This 0.25~phd requirement accepts 98\% of single photoelectrons in each channel, which leads to a S1 detection efficiency determined by the two-fold coincidence requirement. 
The S1 threshold in the signal model required that two individual photons were detected with a summed area above 1.1~phd.
This 1.1~phd requirement was varied to estimate the systematic uncertainty associated with the S1 threshold. 
The single-scatter event distribution after the application of this threshold is shown in Fig.~\ref{fig:final_single_scatter_sim_ly_plots_final_single_scatters_s1_vs_s2_subplots}.
The corresponding $S2$ spectrum is shown by the shaded red histogram in Fig.~\ref{fig:final_single_scatter_sim_ly_plots_final_single_scatters_ly_s2_spectrum}.
Systematic uncertainties associated with this model are discussed in Sec.~\ref{sec:dd_low_energy_ly_background_and_uncertainties}.

\subsection{Data analysis} \label{sec:dd_low_energy_ly_data_analysis}

\begin{figure}[!htbp]
    \begin{center}
        \includegraphics[width=0.48\textwidth]{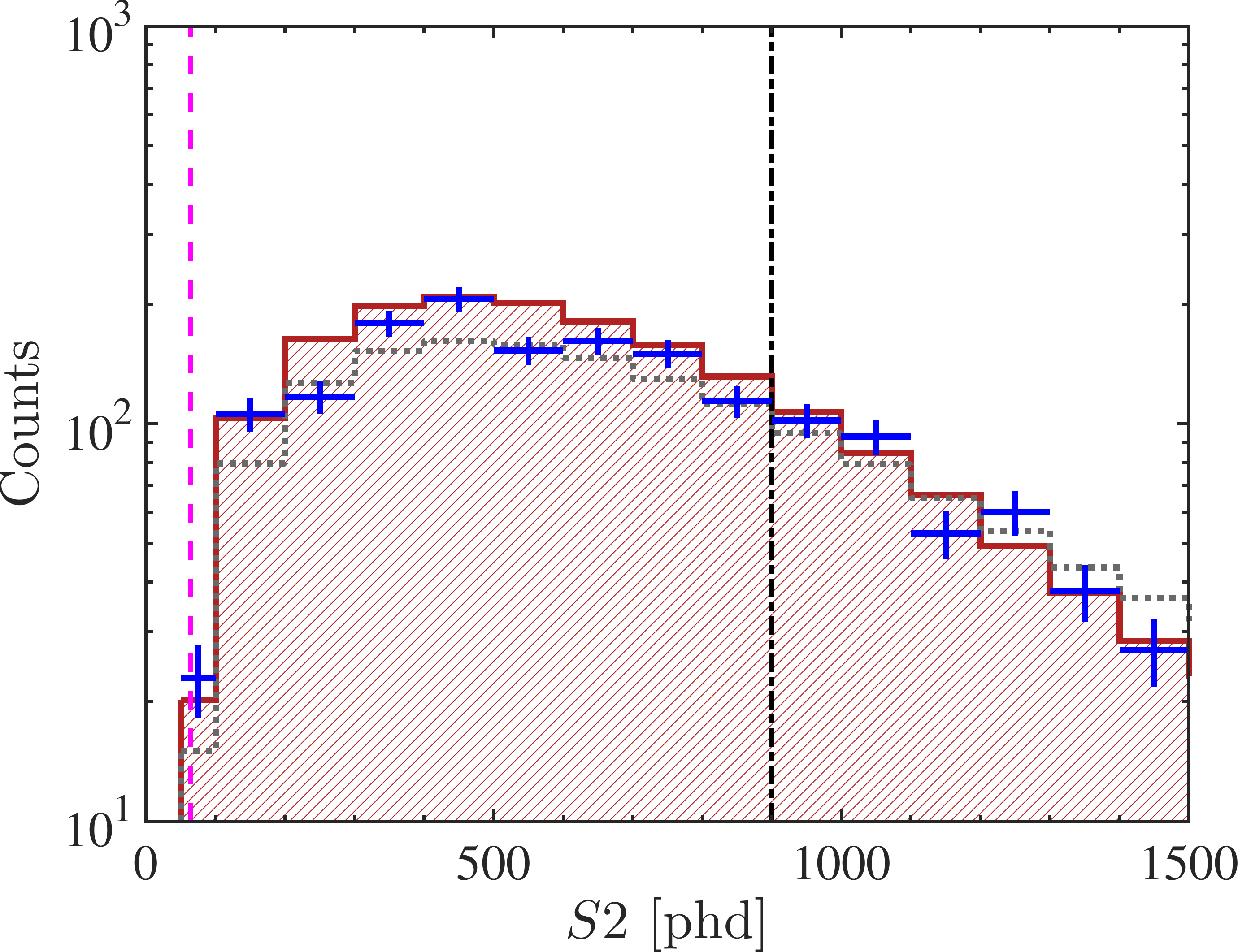}
        \vskip -0.1cm
        \caption{
            The single-scatter $S2$ spectrum from data after all \ly{} analysis cuts are applied is shown in blue.
            The blue error bars are statistical.
            The corresponding simulated $S2$ spectrum is represented by the shaded red histogram; the simulation used the JENDL-4.0 nuclear database.
            The simulated $S2$ spectrum produced using an alternative nuclear database (ENDF/B-VII.1~\cite{ChadwickHermanOblozinskyEtAl2011}) is shown by the gray dotted line.
            The statistical uncertainty on the simulated spectra is negligible.
            The black dot-dashed line at 900 phd $S2$ demarcates the measurement region from the normalization region.
            The simulation is normalized to match the total number of counts observed in data between 900--1500 phd, while the \ly{} points are determined using the events in the region $50 < S2 < 900$.
            The raw $S2 > 55$~phd threshold is represented by the vertical dashed magenta line in corrected $S2$ space ($\sim$64~phd).
        }
        \vskip -0.5cm
        \label{fig:final_single_scatter_sim_ly_plots_final_single_scatters_ly_s2_spectrum}
    \end{center}
\end{figure}

The selected single-scatter events were collected into bins of $S2$, and the resulting mean photon yield for each bin was extracted by comparing the $S1$ distribution of events in the $S2$ bin to the model described in Sec.~\ref{sec:dd_low_energy_ly_signal_model}.
The resulting $S2$ spectrum is shown in Fig.~\ref{fig:final_single_scatter_sim_ly_plots_final_single_scatters_ly_s2_spectrum} for both data and simulation.
The first bin spans the range $50 < S2 < 100$~phd and the subsequent eight bins are 100~phd wide.
The simulation was normalized to match the total number of counts observed in data between 900--1500~phd, while the \ly{} measurement was made using the $S2$ range between 50--900~phd.
This ensured that the normalization between simulation and data was performed outside of the region used to produce \ly{} data points. 
The energy normalization range from 900--1500~phd corresponds to roughly 20--30~\kevnr{}, while the \ly{} measurement region from 50--900~phd contains events with a recoil energy of 0--20~\kevnr{}.
These $S2$ ranges for the \ly{} measurement and normalization regions were chosen to match the recoil energy range used for the forward-scatter based low-energy \qy{} measurement, which accepts events with a maximum recoil energy of 30~\kevnr{}.
The transition between the measurement and normalization regions at an $S2$ of 900~phd was chosen to ensure the measured \ly{} data points fall within the 0.7--24.2~\kevnr{} recoil energy range, where the ionization yield is absolutely defined by the low-energy \qy{} measurement described in Sec.~\ref{sec:dd_low_energy_qy}.
As the \ly{} bin boundaries are defined by $S2$ values, the corresponding mean recoil energy for each \ly{} bin is not expected to match the measured nuclear recoil energy of the \qy{} data points.

The best-fit light yield for each $S2$ bin was obtained via a maximum-likelihood-based optimization of the simulated $S1$ spectrum.
The log-likelihood function is given by

\begin{multline} 
    \label{eq:ly_log_likelihood}
    \ln{L} = \\
    -(N_{s} + N_{b}) + \ln{(N!)} + \ln{\left[\frac{1}{\sqrt{2\pi}\sigma_{g_1}}e^{- \frac{(g_1-g_{1,0})^{2}}{2\sigma_{g_1}^{2}}}\right]} \\
    +\sum\limits_{i=1}^{N}\ln{\left[N_{s} p_{s}(x_{i} \vert \xi, g_1) + N_{b}p_{b}(x_{i}) \right]} \text{,}
\end{multline}

\noindent
where the parameters are $\xi$, $N_{s}$, $N_{b}$, and $g_1$.
The constant $N$ is the total number of events in the $S2$ bin of interest in data.
It does not vary during the optimization so the $\ln{(N!)}$ term can be dropped.
The parameter $\xi$ is a scaling factor used to vary the light yield in the simulation during the optimization.
The parameter $N_{s}$ is the number of signal events expected from simulation based upon the normalization at higher energies.
This value of $N_{s}$ is fixed for each iteration of the $\xi$ parameter. 
The parameter $N_{b}$ is the number of events in a floating flat background PDF component; this is typically $\sim$1\% and no more than $\sim$10\% of total events.
The parameter $g_1$ is the S1 photon detection efficiency, which is allowed to float as a nuisance parameter within the constraints set by the measured value of $0.115 \pm 0.004$.
This treatment of $g_1$ as a constrained nuisance parameter incorporates the systematic uncertainty due to the photon detection efficiency into the reported uncertainties resulting from the four-dimensional log-likelihood contour.

This optimization was performed for each of the nine $S2$ bins used to extract \ly{}.
The resulting parameters of interest are the measured mean number of S1 photons leaving the interaction site, \np{}, and the mean underlying recoil energy. 
After the optimization, these parameters were determined from the model with the best fit to the observed S1 photon distribution.
The corresponding \ly{} data point is centered on the mean energy of the underlying Monte Carlo events that populate the $S2$ bin.

\begin{figure}[!htbp]
    \begin{center}
        \includegraphics[width=0.480\textwidth]{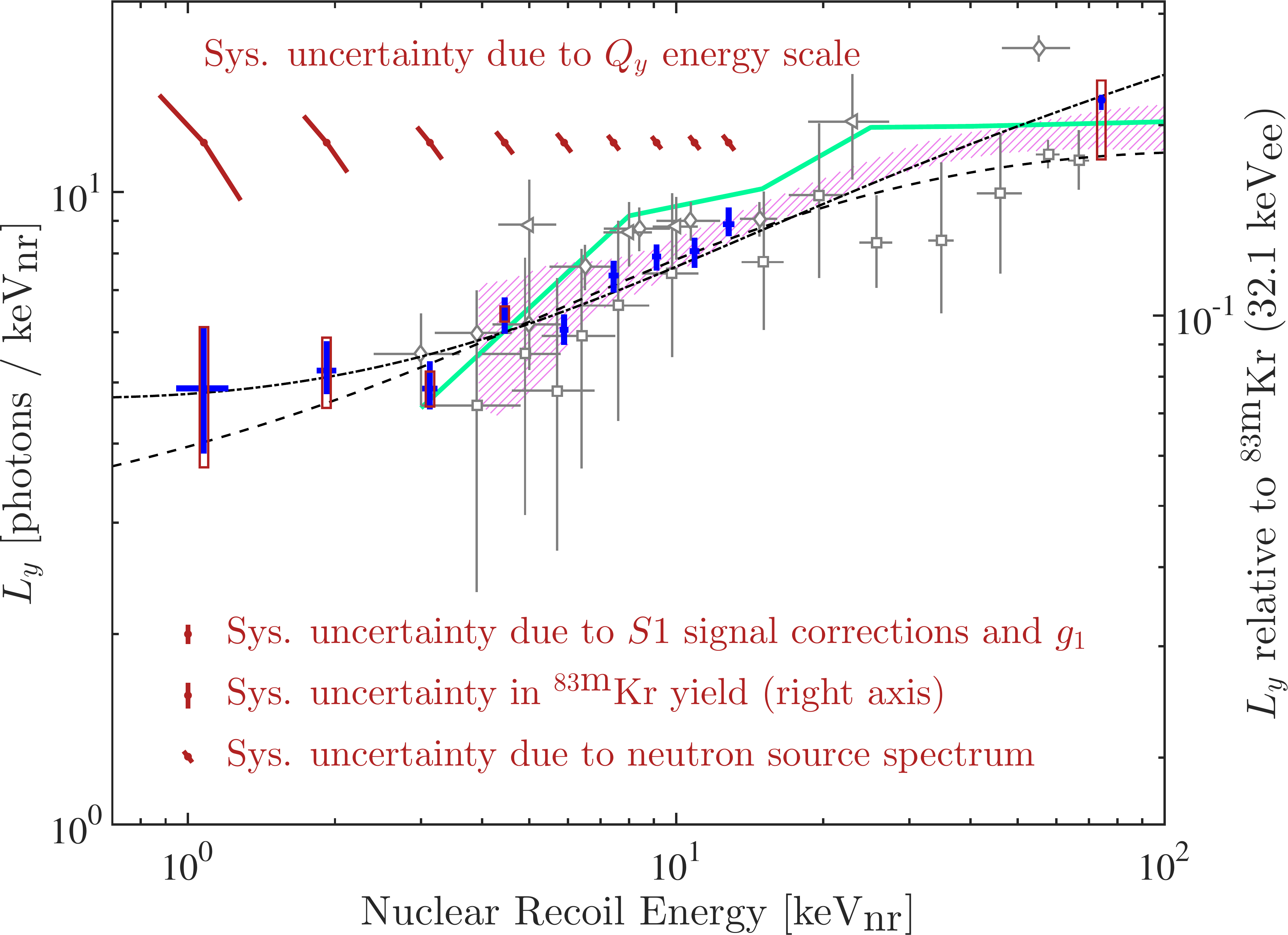}
        \vskip -0.1cm
        \caption{
            The LUX \ly{} measured at 180~V/cm is shown by the blue points.
            The left axis is the absolute yield \ly{} in units of photons/\kevnr{} and the right axis is the \ly{} relative to the LUX \insitu{} 32.1~\kevee{}~$^{83\textrm{m}}$Kr yield at 0~V/cm.
            The red diagonal error bars at the top of the plot correspond to the $1 \sigma$ systematic uncertainties on the determination of the energy scale from our measured \qy{}.
            Inserted below the data, the top red systematic uncertainty marker on the left side of the plot corresponds to the scaling uncertainty due to $g_1$ and $S1$ signal corrections (shown for reference here as $g_1$ was included as a nuisance parameter).
            The middle red systematic uncertainty marker is the uncertainty on the $^{83\textrm{m}}$Kr light yield as measured at 0~V/cm in LUX.
            This uncertainty is applicable to the right axis scale only.
            The bottom red systematic uncertainty marker corresponds to the uncertainty in the mean neutron energy produced by the \dd{} source.
            The red box indicates the systematic uncertainty in the endpoint \ly{} measurement at 74~\kevnr{}.
            The gray data points represent other angle-based measurements with a \kevnr{} energy scale.
            The \leff{} measurements in Refs.~\cite{AprileBaudisChoiEtAl2009} ($\triangleleft$),~\cite{Manzur2010} ($\Box$), and~\cite{Plante2011} ($\Diamond$) were performed at 0~V/cm.
            The purple band~\cite{Horn2011} and thick green line~\cite{AprileAlfonsiArisakaEtAl2013} represent spectral fit based \leff{} measurements with a \kevnr{} energy scale corrected to 0~V/cm using an assumed value of $S_{\textrm{nr}}$.
            The dashed (dot-dashed) black line corresponds to the Lindhard-based (Bezrukov-based) LUX best-fit NEST model described in Sec.~\ref{sec:nest_post_dd}.
        } 
        \vskip -0.5cm
        \label{fig:scintillation_yield_alternate_alpha_scaling_s1_only_light_yield_final}
    \end{center}
\end{figure}

The resulting measured scintillation yield for each of the nine bins is shown in Fig.~\ref{fig:scintillation_yield_alternate_alpha_scaling_s1_only_light_yield_final}.
The LUX \ly{} measurement is shown in both absolute units of photons/\kevnr{} on the left axis and relative to the 32.1~\kevee{}~$^{83\textrm{m}}$Kr light yield as measured at 0~V/cm on the right axis.
It is worth noting that the left axis represents the first direct absolute measurement of the nuclear recoil scintillation yield.
All previous measurements of the liquid xenon light yield have reported results in terms of \leff{}, the observed light yield relative to that of 122~\kevee{} gamma rays from a $^{57}$Co calibration source.

We performed a cross-check of the observed event rate in data and simulation, similar to what was done for the low-energy \qy{} analysis. 
A single-scatter LUXSim/GEANT4 based simulation of single-scatters was performed using the Lindhard-based NEST model described in Sec.~\ref{sec:nest_post_dd}.
The simulation output was used to produce per-channel waveform data that was processed using the LUX \dd{} analysis pipeline.
A neutron source rate of $(2.6 \pm 0.8)~\times~10^6$~n/s provides the optimal match between the absolute number of single-scatter events in simulation and data, which is in agreement with the independently measured neutron production rate of $(2.5 \pm 0.3)~\times~10^{6}$~n/s.
In addition, this absolute rate is consistent with the corresponding normalization of double scatter data in Sec.~\ref{sec:dd_low_energy_qy}.
This agreement between the observed data and expected absolute event rate using the model in Sec.~\ref{sec:nest_post_dd} demonstrates the self-consistency of the measured yields with the observed number of events.
Additionally, the agreement with the independently measured best-fit neutron rate at the source from the double-scatter \qy{} analysis demonstrates consistency between the single-scatter and double-scatter analyses. 

The relative scale (right vertical axis) in Fig.~\ref{fig:scintillation_yield_alternate_alpha_scaling_s1_only_light_yield_final} is set using the measured $^{83\textrm{m}}$Kr yield in LUX at 0~V/cm of $64 \pm 3$ photons/\kevee{}.
The internal, homogeneous $^{83\textrm{m}}$Kr source is a more effective standard candle than the primary gamma ray produced by the external $^{57}$Co source traditionally used for \leff{} measurements due to the self-shielding properties of large liquid xenon TPCs.
$^{83\textrm{m}}$Kr decays via the emission of a 32.1~\kevee{} conversion electron followed by a 9.4~\kevee{} conversion electron with a characteristic time separation of about 154~ns.\footnote{Unlike the 9.4~\kevee{} component, the light yield of the 32.1~\kevee{} component is constant as a function of the time separation between the emission of conversion electrons and can be used as a standard candle~\cite{Baudis2013a}.}
Previous measurements reported in terms of \leff{} were converted to \ly{} assuming a $^{57}$Co absolute yield of 63~photons/\kevee{} at 0~V/cm~\cite{SzydagisFyhrieThorngrenEtAl2013, StephensonHaefnerLinEtAl2015}.
Conveniently in LUX, as was found in Ref.~\cite{ManalaysayUndagoitiaAskinEtAl2010}, the $^{83\textrm{m}}$Kr yield at 32.1~\kevee{} and the $^{57}$Co yield at 122~\kevee{} are in close agreement allowing easy direct comparison to previous \leff{} measurements using the right axis in Fig.~\ref{fig:scintillation_yield_alternate_alpha_scaling_s1_only_light_yield_final}.

The electron recoil light yield was also measured using $^{131\textrm{m}}$Xe remaining in the liquid xenon from cosmogenic activation before the target media was transported underground.
The $^{131\textrm{m}}$Xe nuclei undergoes an isomeric transition depositing 163.9~\kevee{} with a half life of 11.8~days and provides an internal, homogeneous calibration source close in energy to the 122~\kevee{} gamma from $^{57}$Co that has been used to calibrate smaller liquid xenon TPCs in the past.
The light yield for 163.9~\kevee{} electron recoils was measured to be $41.3 \pm 1.1$~photons/\kevee{} at 180~V/cm using the $^{131\textrm{m}}$Xe source.
We can then extrapolate the light yield from this data point to the commonly used standard candle energy of 122~\kevee{} using NEST~v0.98. 
The light yield due to a 122~\kevee{} electron recoil at 180~V/cm is 1.12$^{+0.08}_{-0.06}$~times higher than the yield at 164~\kevee{} according to NEST~v0.98.
After accounting for this yield translation factor and the expected $S_{\textrm{ee}}$($\mathcal{E} = 180$~V/cm) field quenching factor for electron recoils of 0.74~\cite{Aprile2006a, ManalaysayUndagoitiaAskinEtAl2010}, we measure the electron recoil yield for a 122~\kevee{} gamma ray to be $63^{+5}_{-4}$~photons/\kevee{} at 0~V/cm.
This measured light yield for 122~\kevee{} electron recoils in LUX is in agreement with the value of 63~photons/\kevee{} at 0~V/cm used to convert previous \leff{} results to \ly{}.

Avoiding any assumptions about $S_{\textrm{nr}}$ and $S_{\textrm{ee}}$, the LUX measured \ly{} in Fig.~\ref{fig:scintillation_yield_alternate_alpha_scaling_s1_only_light_yield_final} is reported in absolute units at 180~V/cm.
Previous results in the figure were measured at 0~V/cm or were corrected to 0~V/cm assuming various values of $S_{\textrm{nr}}$ for the operating field---all of which ranged from 0.92--1.0.
The agreement of results from liquid xenon TPCs operating across a broad range of drift fields (0--3.6~kV/cm) in Fig.~\ref{fig:scintillation_yield_alternate_alpha_scaling_s1_only_light_yield_final} indicates that the nuclear recoil light yield in liquid xenon is a weak function of the drift electric field.

\subsection{Backgrounds and uncertainties} \label{sec:dd_low_energy_ly_background_and_uncertainties}

Accidental coincidences due to S2 signals produced by delayed extraction of ionization electrons and photoionization of impurities can masquerade as single-scatter events potentially contributing to a background in the lowest-energy \ly{} bin ($50 < S2 < 100$~phd).
Overlapping photoelectrons due to the intrinsic PMT dark rate as well as stray photons contribute to the S1 signals in these accidental coincidence single-scatter events.
These accidental coincidence events are rejected using a combination of two data quality cuts: the upper limit on the raw pulse area in the event window outside of the S1 and S2 signals, and the requirement that there are no single electrons or S2 signals before the S1 in the event record.

The number of accidental coincidence events remaining after the application of the two data quality cuts was quantified using off-beam single-scatter interactions as a sideband.
The accidental coincidence events potentially contributing to a low-energy \ly{} analysis background were verified to occur with a flat distribution as a function of $z$. 
The background analysis volume was offset to $z=33.9$~cm below the liquid surface, away from the true neutron beam center at $z=16.1$~cm.
Other than the analysis volume offset, an identical analysis was performed. 
Any events observed in the first $S2$ bin of the sideband analysis were conservatively assumed to be accidental coincidences.
This conservative estimate of the accidental coincidence event contamination in the first \ly{} bin is $3.0 \pm 1.7$ events, which is within the $1\sigma$ uncertainty of the total number of events in this bin during the standard analysis.

The \ly{} data and per-bin statistical and systematic uncertainties are listed in Table~\ref{tab:dd_ly_result}.
Uncertainties common across all bins in the low-energy and endpoint \ly{} measurements are listed in Table~\ref{tab:dd_common_ly_uncertantites}.

\begin{table}[!htbp]
    \centering
    \caption{
        The measured scintillation yield for nuclear recoils in liquid xenon at an electric field of 180~V/cm and associated statistical uncertainties.
        The first two columns correspond to the blue low-energy \ly{} data points in Fig.~\ref{fig:scintillation_yield_alternate_alpha_scaling_s1_only_light_yield_final}.
        The fractional systematic uncertainty in energy due to the data-driven \qy{} based energy scale is denoted by $\Delta E_{\textrm{nr}} / E_{\textrm{nr}}$.
        This uncertainty in energy is represented by the slanted red error bars at the top of the figure due to the anti-correlation of the location of the \ly{} data points on the vertical axis. 
        The estimated fractional systematic uncertainty in \np{} due to uncertainty in the S1 threshold is represented by $\Delta$\np{}/\np{}.
        This uncertainty is represented by the red boxes around the low-energy \ly{} data points.
        Quoted uncertainties are symmetric ($\pm$) unless otherwise indicated.
    }
    \label{tab:dd_ly_result}
    \setlength{\extrarowheight}{.5em}
    \begin{tabular*}{\columnwidth}{@{\extracolsep{\fill}} SSSS}
        \hline \hline
        {$E_{\textrm{nr}}$} & {\ly{}} & {$\Delta E_{\textrm{nr}} / E_{\textrm{nr}}$} & {$\Delta$\np{}/\np{}} \\
        {(\kevnr{})} & {(ph/\kevnr{})} & {(\%)} & {(\%)} \\
        \hline
        1.08~$\pm$~0.13 & 4.9 $^{+1.2}_{-1.0}$ & 19 & 25 \\
        1.92~$\pm$~0.09 & 5.2 $^{+0.6}_{-0.4}$ & 10 & 13 \\
        3.13~$\pm$~0.11 & 4.9 $^{+0.5}_{-0.4}$ & 6 & 6 \\
        4.45~$\pm$~0.11 & 6.4 $^{+0.4}_{-0.4}$ & 4 & 3 \\
        5.89~$\pm$~0.13 & 6.1 $^{+0.4}_{-0.3}$ & 3 & {-} \\
        7.44~$\pm$~0.17 & 7.4 $^{+0.4}_{-0.4}$ & 3 & {-} \\
        9.1~$\pm$~0.2 & 7.9 $^{+0.4}_{-0.4}$ & 3 & {-} \\
        10.9~$\pm$~0.3 & 8.1 $^{+0.4}_{-0.5}$ & 2 & {-} \\
        12.8~$\pm$~0.3 & 8.9 $^{+0.6}_{-0.4}$ & 3 & {-} \\
        \hline \hline
    \end{tabular*}
\end{table}

The systematic uncertainty in \np{} (and \ly{} as it is proportional to \np{}) due to the S1 threshold model used in simulation is reported in Table~\ref{tab:dd_ly_result}.
The contribution from the S1 threshold model to the systematic uncertainty in \ly{} was estimated by re-analyzing the data using the alternative S1 thresholds in the signal model, as indicated by the arrows in Fig.~\ref{fig:final_single_scatter_sim_ly_plots_final_single_scatters_s1_vs_s2_subplots}.
The two alternative S1 thresholds require at least two photons to be detected in simulation with a combined area of 0.5~phd and 2.0~phd, respectively.
The systematic uncertainty due to the modeled S1 threshold was conservatively estimated by quoting the maximum variation in \ly{} observed during this exercise.
The average systematic uncertainty due to the measurement uncertainty in $g_1$ is quantified in Table~\ref{tab:dd_common_ly_uncertantites}.

The dominant effect contributing to the $S1$ resolution is Poisson fluctuation in the number of collected photons. 
Resolution effects due to variations in the underlying Fano factor and recombination fluctuations were confirmed to be subdominant.
Systematic effects due to S2 threshold uncertainty, which only affect the lowest recoil energy bin, were confirmed to be subdominant to the reported uncertainties for 10\% variations in S2 threshold.

If the nuclear database used in the signal model overpredicts (underpredicts) the expected number of events at low energies the optimization will favor a lower (higher) \ly{} to compensate. 
The JENDL-4.0 library used in the model used to extract the \ly{} result is the most modern evaluation for neutron-xenon cross-sections for neutron energies of 2.45~MeV~\cite{Robinson2014}.
Seven other nuclear databases were studied to quantify the effect on the predicted number of events at low energies and the effect on the measured \ly{} when using older evaluations.
Of the databases studied, ENDF/B-VII.1 and JENDL-4.0 represent the extremes in the angular scattering cross-section over the energy range of interest for this analysis between 0--30~\kevnr{} (roughly $S2 < 1500$~phd).
In addition to being the most modern evaluation, the baseline JENDL-4.0 database is the most conservative for use in the light yield measurement as all other databases predict fewer events at low energies after normalization between 900--1500~phd $S2$.
The $S2$ spectra produced via the signal model described in Sec.~\ref{sec:nest_post_dd} using both the ENDF/B-VII.1 and JENDL-4.0 databases are shown in Fig.~\ref{fig:final_single_scatter_sim_ly_plots_final_single_scatters_ly_s2_spectrum}.
The \ly{} analysis was repeated using the alternative ENDF/B-VII.1 database, which results in a measured \ly{} 25\% larger at 1.1~\kevnr{}.
The difference in measured \ly{} between databases decreases with increasing energy until it is subdominant to statistical uncertainties at 5.9~\kevnr{}.

As a cross-check, the \ly{} measurement was repeated using an alternative initial modeled light yield in simulation as the starting point for the optimization.
The results of this cross-check were consistent with the baseline measurement within $1\sigma$ statistical uncertainties.

\begin{table}[!htbp]
    \centering
    \caption{
        Uncertainties common to the \ly{} measurement both at low energies and at the \dd{} recoil spectrum endpoint.
        The second column lists systematic uncertainties associated with the mean reconstructed number of primary scintillation photons, \np{}.
        The third column lists systematic uncertainties associated with the mean reconstructed energy, $E_{\textrm{nr}}$.
        Quoted uncertainties are symmetric ($\pm$) unless otherwise indicated.
    }
    \label{tab:dd_common_ly_uncertantites}
    \setlength{\extrarowheight}{.5em}
    \begin{tabular*}{\columnwidth}{@{\extracolsep{\fill}} lcc}
        \hline \hline
        {Source of Uncertainty} & {$\Delta$\np{}/\np{}} & {$\Delta E_{\textrm{nr}} / E_{\textrm{nr}}$} \\
        & {(\%)} & {(\%)} \\
        \hline
        $g_1$ & 3.48 & - \\
        $S1$ correction (3D position)  & 0.6 & - \\
        $S1$ correction (non-uniform field) & $^{+0}_{-0.5}$ & - \\
        Mean neutron energy from source & - & 2 \\
        \hline
        Total & $^{+4}_{-4}$& 2 \\
        \hline \hline
    \end{tabular*}
\end{table}

\section{Signal yields at the D-D neutron recoil spectrum endpoint} \label{sec:dd_endpoint}

The maximum nuclear recoil energy produced in liquid xenon by the mono-energetic 2.45~MeV neutrons from the \dd{} source is given by Eq.~\ref{eq:recoil_energy_equation}.
This provides a known endpoint feature in the $S1$ and $S2$ spectra produced by 180$^{\circ}$ scatters corresponding to the maximum recoil energy of 74~\kevnr{}. 
The population of single-scatter events was used to extract \ly{} and \qy{} using the nuclear recoil energy spectrum endpoint closely following the analysis procedure used in Ref.~\cite{JoshiSangiorgioBernsteinEtAl2014}.
As in the low-energy \ly{} analysis, the neutron beam energy purity cuts and a radial position cut of r~$<$~21~cm were applied.
A raw $S2$ analysis threshold of 164~phd was applied, well below the region of interest near the endpoint where the mean $S2$ is 2500~phd.
An upper limit on the total digitized area in the event record outside of the identified S1 and S2, identical to the one used for the \ly{} analysis, was applied.
As in the previous analyses, an upper limit on the S2 signal RMS width of 775~ns was enforced.
As in the low-energy \ly{} measurement, we used the JENDL-4.0 nuclear database to generate the underlying nuclear recoil energy spectrum in the model used for parameter optimization.

\subsection{Scintillation yield at the nuclear recoil energy spectrum endpoint}

To extract the light yield at 74~\kevnr{}, we fit an $S1$ signal model to the $S1$ spectrum endpoint feature.
The observed $S1$ spectrum was modeled using a constant \ly{} across the entire nuclear recoil energy range.
Three parameters were varied in a binned maximum-likelihood optimization: the \ly{} value at the endpoint as the target parameter, $F_{1}^{\prime}$ as a resolution term, and the overall normalization of counts in the model.
The $S1$ resolution as a function of the mean integer number of photons detected, $n_{\textrm{phd}}$, was set by 

\begin{equation} \label{eq:dd_endpoint_s1_resolution}
    \sigma_{S1}(n_{\textrm{phd}}) = \sqrt{n_{\textrm{phd}}(F_{1}^{\prime} + \sigma^{2}_{\textrm{sphe}})} \,\text{,}
\end{equation}

\noindent
where $\sigma_{\textrm{sphe}} = 0.37$ is the mean single photoelectron resolution of the LUX PMTs~\cite{AkeribBaiBedikianEtAl2012a}.
The $F_{1}^{\prime}$ parameter was allowed to float as a nuisance parameter controlling effective $S1$ resolution in the optimization and accommodating fluctuations in the observed signal.
The most significant contribution to the $F_{1}^{\prime}$ resolution term is the detector's scintillation photon detection efficiency ($g_1$).
The fluctuations associated with scintillation and recombination at the interaction site are subdominant.

\begin{figure}[!htbp]
    \begin{center}
        \includegraphics[width=0.48\textwidth]{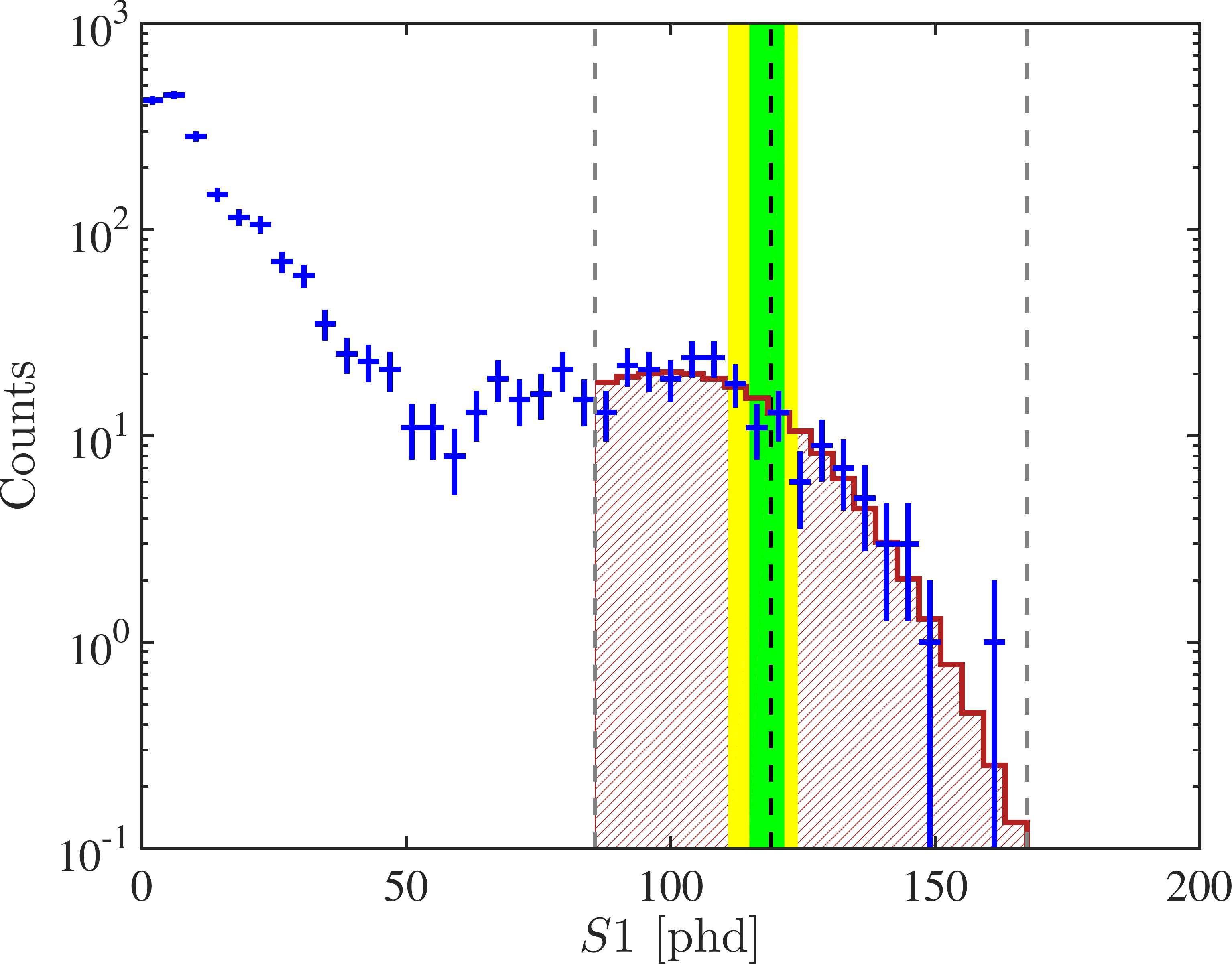}
        \vskip -0.1cm
        \caption{
            Result of the \ly{} endpoint optimization.
            The single-scatter $S1$ spectrum after all cuts is shown in blue.
            The dip in the spectrum at ${\sim} 50$~phd is due to the differential cross-section for elastic neutron scatters on xenon. 
            The horizontal error bar represents the bin width, and the vertical error bar represents the statistical uncertainty on the number of events in each bin.
            The best-fit endpoint model is represented by the red shaded histogram.
            The binned maximum-likelihood optimization was performed between the gray dashed lines.
            The fit has a $\chi^{2}/\text{dof} = 7.5 / 9$ yielding a p-value of 0.59.
            The black dashed line is the best-fit endpoint in $S1$ space, with 1$\sigma$ and 2$\sigma$ statistical uncertainties represented by the green and yellow regions, respectively.}
        \vskip -0.5cm
        \label{fig:endpoint_yield_ly_endpoint_spectrum_with_reconstructed_endpoint_and_fit}
    \end{center}
\end{figure} 

The results of the \ly{} measurement using the nuclear recoil spectrum endpoint are shown in Fig.~\ref{fig:endpoint_yield_ly_endpoint_spectrum_with_reconstructed_endpoint_and_fit}.
The \ly{} at 74~\kevnr{} was measured to be $14.0^{+0.3 \text{(stat)} + 1.1 \text{(sys)}}_{-0.5 \text{(stat)} -2.7 \text{(sys)}}$ photons/\kevnr{} at 180~V/cm.
The systematic uncertainties specific to the \ly{} measurement at the \dd{} recoil spectrum endpoint are listed in the right column of Table~\ref{tab:dd_endpoint_ly_uncertantites} and are represented by the red box around the blue endpoint in Fig.~\ref{fig:scintillation_yield_alternate_alpha_scaling_s1_only_light_yield_final}.
Additional sources of systematic uncertainty that are common to both the endpoint and low-energy \ly{} measurement are listed in Table~\ref{tab:dd_common_ly_uncertantites}.

The \ly{} endpoint specific systematic uncertainties were determined by varying the associated model or analysis parameters and re-running the optimization.
The systematic uncertainty due to the \dd{} neutron recoil energy spectrum used in the model was conservatively estimated by repeating the analysis assuming a completely flat recoil spectrum extending to 74~\kevnr{}.
The systematic uncertainty due to the assumption of a constant \ly{} was determined by repeating the analysis while modeling the \ly{} as a straight line.
In this case, the slope of the modeled \ly{} was allowed to float as an additional nuisance parameter.
The systematic uncertainties due to the choice of optimization region and bin size were estimated by separately repeating the analysis with a 20\% larger optimization region and $\times2$ the number of bins, respectively.

\begin{table}[!htbp]
    \centering
    \caption{
        Uncertainties specific to the \ly{} measurement using the \dd{} nuclear recoil spectrum endpoint.
        Quoted uncertainties are symmetric ($\pm$) unless otherwise indicated.
    }
    \label{tab:dd_endpoint_ly_uncertantites}
    \setlength{\extrarowheight}{.5em}
    \begin{tabular*}{\columnwidth}{@{\extracolsep{\fill}} lcc}
        \hline \hline
        {Source of Uncertainty} & {Statistical} & {Systematic} \\
        & {(\%)} & {(\%)} \\
        \hline
        Binned likelihood optimization & $^{+2}_{-3}$ & - \\
        Input recoil energy spectrum & - & 6  \\
        Slope of \ly{} in model & - & $^{+5}_{-18}$ \\
        Choice of optimization region & - & 1.8 \\
        Choice of bin size & - & 0.4 \\
        \hline
        Total & $^{+2}_{-3}$ & $^{+8}_{-19}$ \\
        \hline \hline
    \end{tabular*}
\end{table}

\subsection{Ionization yield at the nuclear recoil energy spectrum endpoint}

An identical procedure to that used for the \ly{} endpoint was used to extract \qy{} using the same population of single-scatter events.
The observed $S2$ spectrum was modeled using a flat \qy{} across the entire \dd{} recoil energy range.
Three parameters were varied in a binned maximum-likelihood optimization: the \qy{} value at the endpoint as the target parameter, $F_{2}^{\prime}$ as a resolution term, and the overall normalization of counts in the model.
The $S2$ resolution as a function of mean integer number of electrons extracted, $n_{e_{S2}}$, was determined by 

\begin{equation} \label{eq:dd_endpoint_s2_resolution}
    \sigma_{\textrm{S2}}(n_{e_{\textrm{S2}}}) = \sqrt{n_{e_{\textrm{S2}}}(\mu^{2}_{\textrm{SE}}F_{2}^{\prime} + \sigma^{2}_{\textrm{SE}})} \,\text{,} 
\end{equation}

\noindent
where $\mu_{\textrm{SE}}$ and $\sigma_{\textrm{SE}}$ are the mean and standard deviation, respectively, of the observed SE distribution in the \dd{} dataset.

The $F_{2}^{\prime}$ parameter was allowed to float as a nuisance parameter controlling effective $S2$ resolution in the optimization and accommodating fluctuations in the observed signal.
The most significant contributions to the $F_{2}^{\prime}$ resolution term are the fluctuations associated with ionization and recombination at the interaction site, as well as binomial detector effects due to the free electron lifetime and electron extraction efficiency.

\begin{figure}[!htbp]
    \begin{center}
        \includegraphics[width=0.48\textwidth]{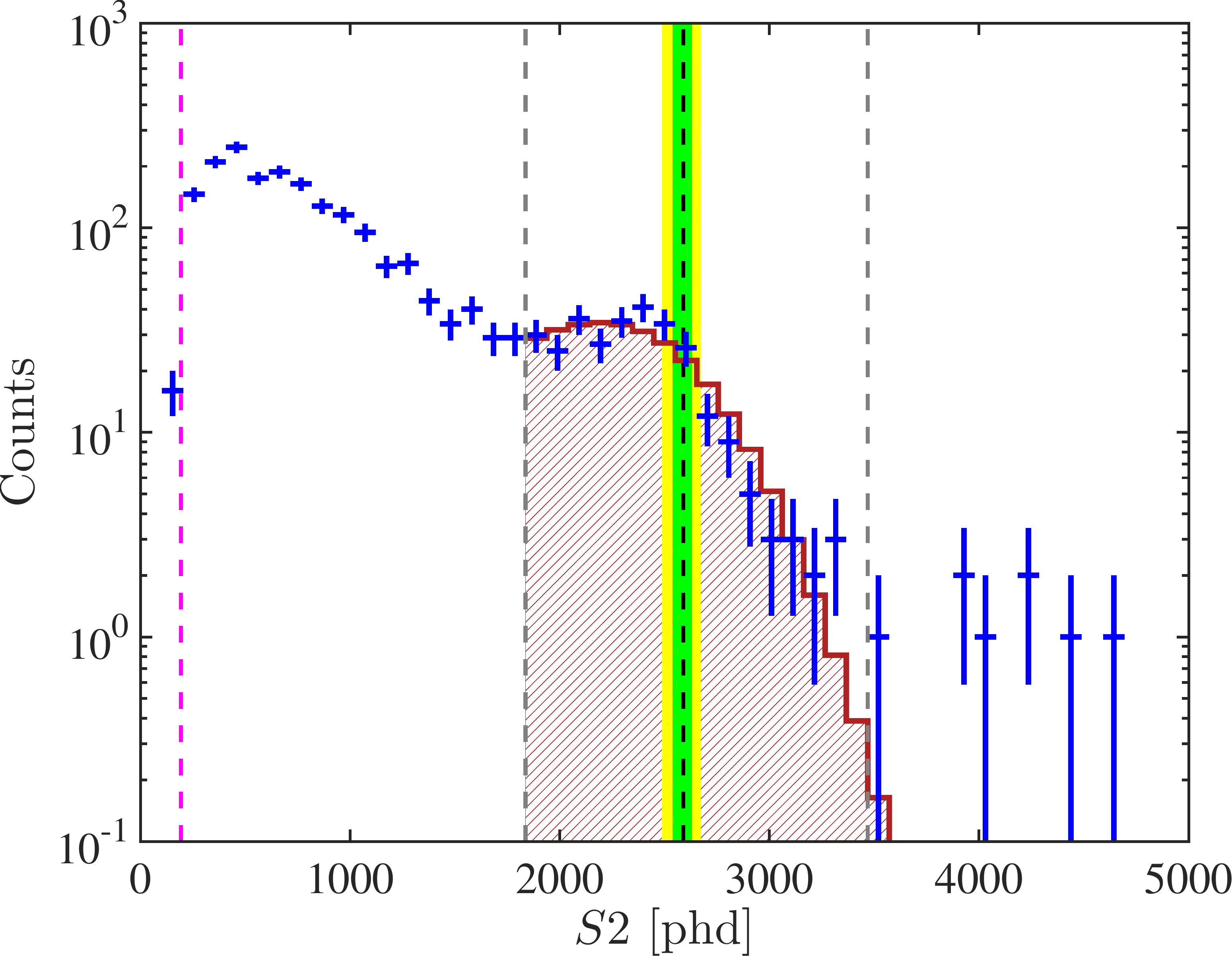}
        \vskip -0.1cm
        \caption{
            Result of the \qy{} endpoint optimization.
            The single-scatter $S2$ spectrum after all cuts is shown in blue.
            The dip in the spectrum at ${\sim} 1750$~phd is due to the differential cross-section for elastic neutron scatters on xenon. 
            The horizontal error bar represents the bin width, and the vertical error bar represents the statistical uncertainty on the number of events in each bin.
            The best-fit endpoint model is represented by the red shaded histogram.
            The binned maximum-likelihood optimization was performed between the gray dashed lines.
            The magenta dashed line depicts the location of the S2 threshold.
            The $\chi^{2}/\text{dof}$ is 14.7/9 dof yielding a p-value of 0.10.
            The black dashed line is the best-fit endpoint in S2 space, with 1$\sigma$ and 2$\sigma$ statistical uncertainties represented by the green and yellow regions, respectively.
            The six events observed outside of the fit range ($3500 < S2 < 5000$) could be due to multiple simultaneous S2 signals at the same $z$ position combined in the event record, residual $^{83\textrm{m}}$Kr from the frequent injections for standard detector corrections, or contamination consisting of 39.6~\kevee{} gamma rays from $^{129}$Xe inelastic neutron scatters.
        }
        \vskip -0.5cm
        \label{fig:endpoint_yield_qy_endpoint_spectrum_with_reconstructed_endpoint_and_fit}
    \end{center}
\end{figure} 

The results of the \qy{} measurement using the \dd{} recoil spectrum endpoint are shown in Fig.~\ref{fig:endpoint_yield_qy_endpoint_spectrum_with_reconstructed_endpoint_and_fit}.
The \qy{} value at 74~\kevnr{} was measured to be $3.06^{+0.05\text{(stat)} + 0.2 \text{(sys)}}_{-0.06 \text{(stat)} -0.4 \text{(sys)}}$ electrons/\kevnr{} at 180~V/cm.
The systematic uncertainties specific to the \qy{} measurement at the nuclear recoil spectrum endpoint are listed in Table~\ref{tab:dd_endpoint_qy_uncertantites} and are represented by the red box around the blue endpoint in Fig.~\ref{fig:ionization_yield_qy_endpoint_ionization_yield}.
Identically to the procedure used for \ly{}, the endpoint-specific systematic uncertainties for \qy{} were determined by varying the associated model or analysis parameters and re-running the optimization.
Additional sources of systematic uncertainty that are common to both the endpoint and low-energy \qy{} measurement are listed in Table~\ref{tab:dd_common_qy_uncertantites}.

\begin{table}[!htbp]
    \centering
    \caption{
        Uncertainties specific to the \qy{} measurement using the \dd{} recoil spectrum endpoint.
        Quoted uncertainties are symmetric ($\pm$) unless otherwise indicated.
    }
    \label{tab:dd_endpoint_qy_uncertantites}
    \setlength{\extrarowheight}{.5em}
    \begin{tabular*}{\columnwidth}{@{\extracolsep{\fill}} lcc}
        \hline \hline
        {Source of Uncertainty} & {Statistical} & {Systematic} \\
        & {(\%)} & {(\%)} \\
        \hline
        Binned likelihood optimization & $^{+1.6}_{-2}$ & - \\
        Input recoil energy spectrum & - & 5  \\
        Slope of \qy{} in model & - & $^{+0.16}_{-11}$ \\
        Choice of optimization region & - & 6 \\
        Choice of bin size & - & 1.6 \\
        \hline
        Total & $^{+1.6}_{-2}$ & $^{+7}_{-13}$ \\
        \hline \hline 
    \end{tabular*}
\end{table}

\section{Nuclear recoil band} \label{sec:dd_nr_band}

The ratio of the ionization to scintillation signal is used to discriminate between nuclear and electron recoils in liquid xenon TPCs.
The band created by nuclear recoil events in $\log_{10}{(S2/S1)}$ vs. $S1$ space is commonly referred to as the ``nuclear recoil band.'' 
In this section, we use neutrons from the \dd{} source to define the nuclear recoil band over the $S1$ range used for the WIMP search analysis.
Subsequently, a simulated nuclear recoil band is compared to data to demonstrate consistency of the nuclear recoil signal model used to generate $S1$ and $S2$ PDFs for the WIMP search profile likelihood ratio analysis~\cite{AkeribAraujoBaiEtAl2015, AkeribAraujoBaiEtAl2016}.

The nuclear recoil band was measured using the single-scatter event population in the \dd{} calibration dataset. 
An S2 threshold at 164~phd was applied on the raw $S2$ area before position-correction for consistency with the LUX WIMP search~\cite{AkeribAraujoBaiEtAl2015}.
An upper limit ensuring $S2 < 5000$~phd was applied.
The nuclear recoil band analysis applied an upper limit on the raw digitized area outside of the identified S1 and S2 signals in the event record of 219~phd.
This cut ensures quiet detector conditions as described in Sec.~\ref{sec:dd_low_energy_ly}.
In contrast to the signal yield measurements presented earlier, the kinetic energy of each incident neutron does not need to be precisely known for the nuclear recoil band measurement.
The neutron beam energy purity cuts (beam line analysis volume) were not applied in order to increase the useful number of neutron events.
Instead, a $z$ cut of $80 < \text{drift time} < 130$~$\mu$s was applied to select events in the plane of the neutron beam projection in the TPC active region.
A radial cut of $r < 21$~cm was used.

\begin{figure}[!htbp]
    \begin{center}
        \includegraphics[width=0.48\textwidth]{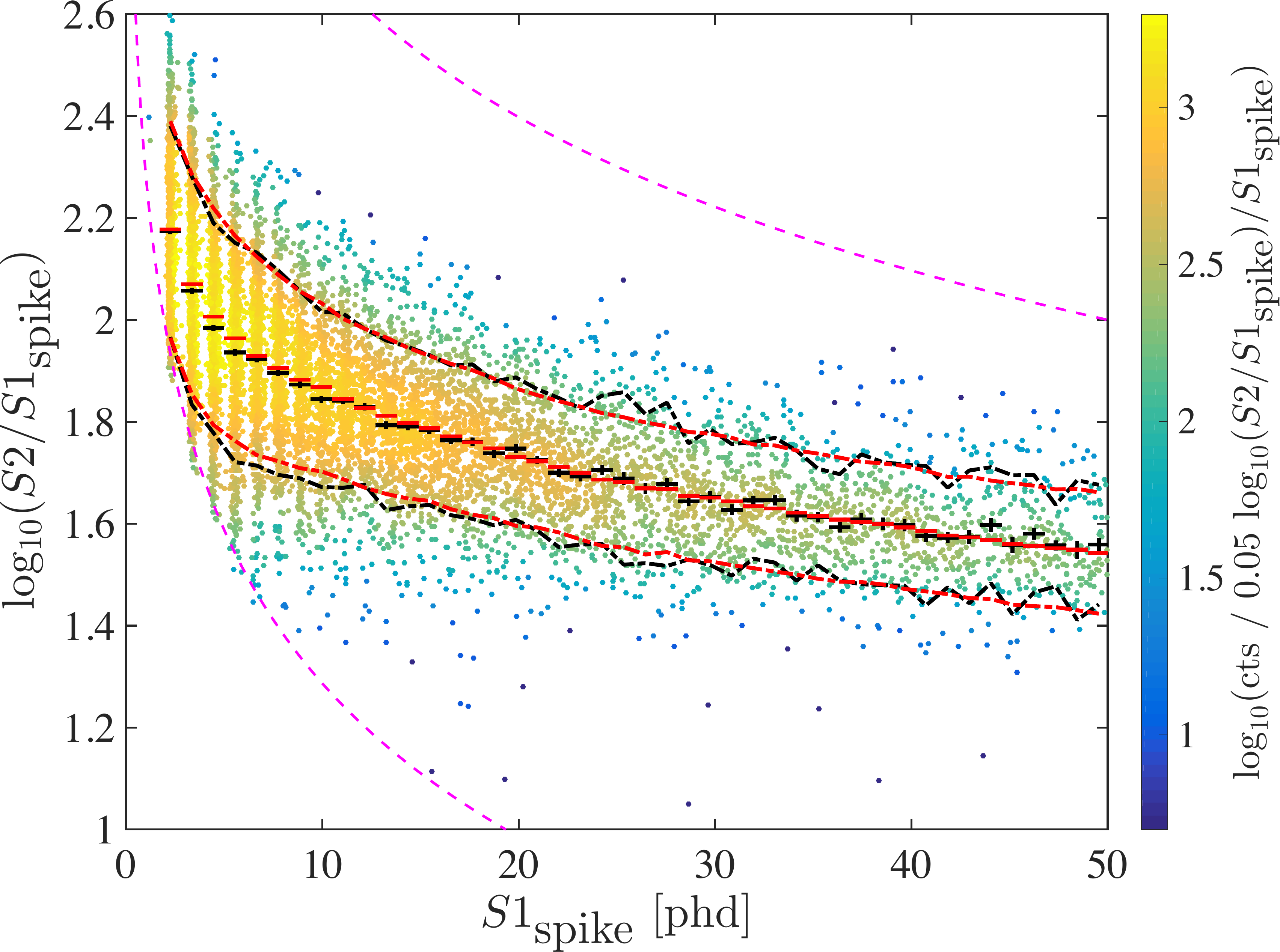}
        \vskip -0.1cm
        \caption{
            The measured events defining the nuclear recoil band are shown in the scatter plot.
            There are 9864 events remaining after all cuts with $S1_{\textrm{spike}} < 50$.
            The black data points are the Gaussian fit centroid values for each $S1_{\textrm{spike}}$ bin.
            The red data points are corresponding Gaussian fit mean value for the simulated nuclear recoil band produced using the model described in Sec.~\ref{sec:nest_post_dd}.
            The black and red dot-dashed lines indicate the 90\% one-sided limits from data and simulation, respectively.
            The magenta dashed lines indicate the lower S2 threshold at $\sim$164~phd raw $S2$ and the upper $S2$ limit at 5000~phd.
            Error bars are statistical only.
        }
        \vskip -0.5cm
        \label{fig:nr_band_luxsim_spike_count_phd_nr_band_ws_s1_range_data_sim_comparison}
    \end{center}
\end{figure}

After all cuts, the remaining events with $S1_{\textrm{spike}} < 50$~phd are shown in Fig.~\ref{fig:nr_band_luxsim_spike_count_phd_nr_band_ws_s1_range_data_sim_comparison}.
This is the same $S1_{\textrm{spike}}$ range used for the improved LUX WIMP search result~\cite{AkeribAraujoBaiEtAl2015}.
The non-zero width of the vertical bands of events at low $S1_{\textrm{spike}}$ is due to corrections for spike overlap in the per-channel waveforms as well as 3D position-based detector corrections.
The mean $S1_{\textrm{spike}}$ value is offset slightly from integer values due to the same corrections.

A Gaussian was fit to the $\log_{10}{(S2/S1_{\textrm{spike}})}$ distribution in each 1.1~phd-wide bin along the $S1_{\textrm{spike}}$ axis.
The Gaussian centroid and 90\% one-sided limit for each bin, depicted in black in Fig.~\ref{fig:nr_band_luxsim_spike_count_phd_nr_band_ws_s1_range_data_sim_comparison}, were determined based upon the fit parameters.
The bins were positioned to ensure the observed vertical bands of events at low $S1_{\textrm{spike}}$ were centered in their corresponding bin.
It is worth noting the significant improvement in the single detected photon resolution at low $S1_{\textrm{spike}}$ compared to traditional $S1$ area-based techniques, which are subject to the intrinsic single photoelectron resolution ($\sigma_{\textrm{sphe}} = 0.37$ in the case of LUX).

In dual-phase liquid xenon TPCs, multiple-scatter events misidentified as single-scatters due to interactions in the reverse field region below the cathode produce events at artificially low $\log_{10}{(S2/S1_{\textrm{spike}})}$~\cite{Angle2008, LebedenkoAraujoBarnesEtAl2009}. 
Compared to more traditional nuclear recoil band calibrations using $^{252}$Cf or $^{241}$Am/Be, there is a relative absence of these pathological events at low $\log_{10}{(S2/S1_{\textrm{spike}})}$ due to the well-defined neutron beam position near the liquid xenon surface away from the sub-cathode ionization signal dead region.

A LUXSim/GEANT4-based simulation using a Lindhard-based NEST model fit to the LUX \dd{} results (described in Sec.~\ref{sec:nest_post_dd}) was used to produce single-scatter event waveforms for comparison with the measured nuclear recoil band.
These waveforms were passed through the data processing pipeline used for the \dd{} calibration data.
The same cuts and analysis procedures used for nuclear recoil band data were applied to the resulting reduced simulation waveforms.
The average (maximum) deviation of the band fit centroid between simulation and data is 0.010 (0.029) in $\log_{10}{(S2/S1_{\textrm{spike}})}$ space over the 0--50~phd $S1_{\textrm{spike}}$ range.
The average standard deviation of the band agrees with a mean (maximum) absolute deviation of 0.009 (0.039) in $\log_{10}{(S2/S1_{\textrm{spike}})}$ space over the 0--50~phd $S1_{\textrm{spike}}$ range.
The simulated nuclear recoil band is consistent with \dd{} calibration data within the systematic uncertainty intrinsic to the simulation process.
This simultaneous agreement of the model described in Sec.~\ref{sec:nest_post_dd} with the measured \qy{}, \ly{}, and nuclear recoil band demonstrates the consistency of the signal model used to generate the WIMP search limit with data~\cite{AkeribAraujoBaiEtAl2015, AkeribAraujoBaiEtAl2016}.
As an additional check, we verified the LUX WIMP search limit is unchanged for all reported WIMP masses by the small variation in the nuclear recoil band between data and simulation.

\begin{figure}[!htbp]
    \begin{center}
        \includegraphics[width=0.48\textwidth]{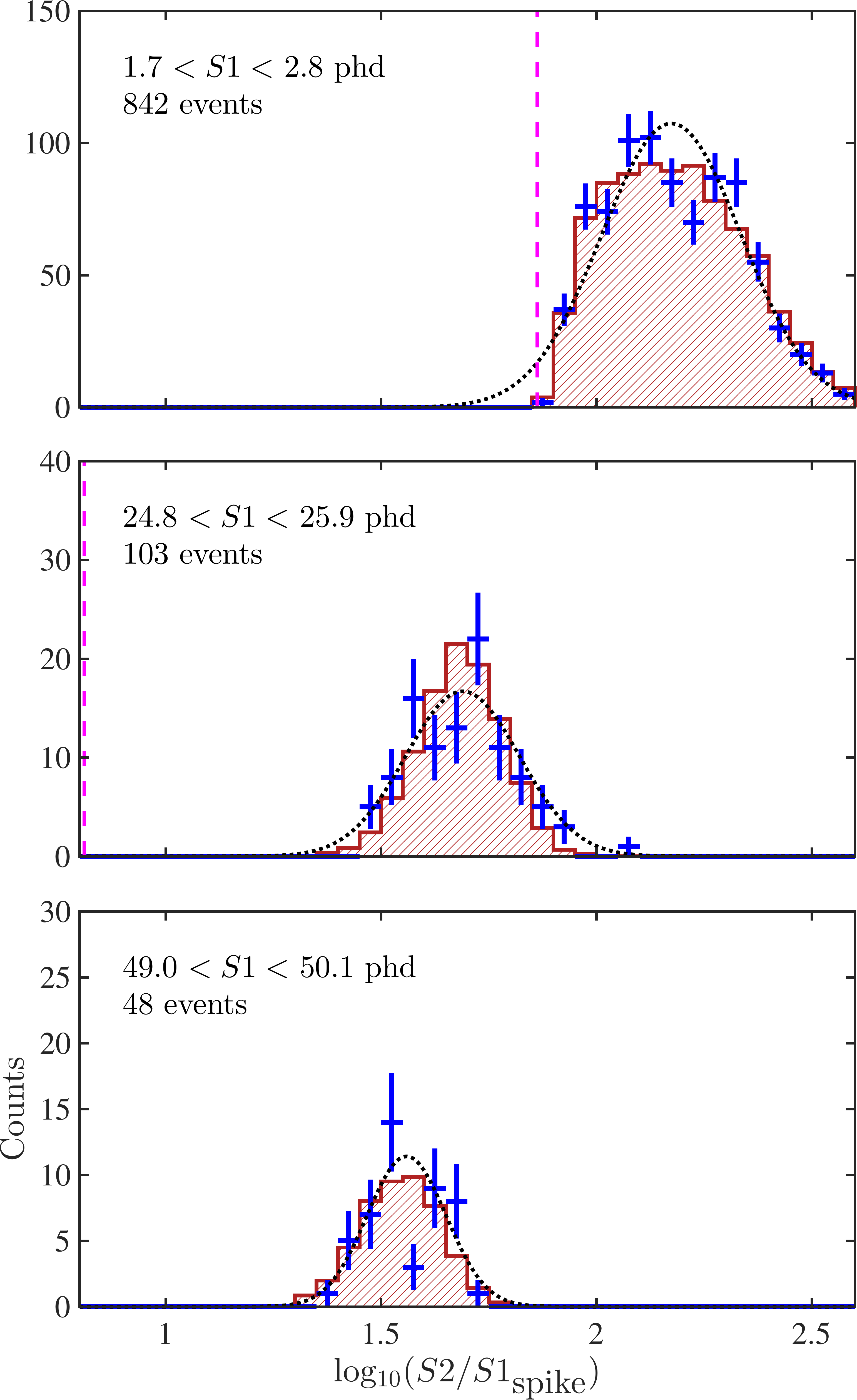}
        \vskip -0.1cm
        \caption{
            Comparison of representative bins used for nuclear recoil band comparison between data and simulation. 
            The lowest, middle, and highest bins in $S1_{\textrm{spike}}$ from Fig.~\ref{fig:nr_band_luxsim_spike_count_phd_nr_band_ws_s1_range_data_sim_comparison} are shown.
            The blue crosses show the distribution of events in data with associated statistical uncertainties.
            The black dotted line shows a Gaussian fit to the blue data points.
            The red shaded histogram represents simulated nuclear recoil band profile.
            The simulation histograms were generated using 9324~events, 3067~events, and 1924~events, respectively, in the three graphs, and the amplitude of each was independently scaled to match the number of events in data. 
            This corresponds to a statistical uncertainty on the maximum value in the red histogram in each graph of 3\%, 4\%, and 5\%, respectively.
            As expected, non-Gaussian behavior is observed in the first $S1_{\textrm{spike}}$ bin.
            The magenta dashed line indicates the approximate location of the S2 threshold.
            The single bin with three counts in the bottom frame is not statistically unreasonable; the $\chi^{2}$/dof for the Gaussian fit in that bin is 9.9/5, which gives a p-value of 0.08.
        }
        \vskip -0.5cm
        \label{fig:nr_band_luxsim_spike_count_phd_representative_bin_sim_comparison}
    \end{center}
\end{figure} 

Representative individual Gaussian fits to data for the lowest, middle, and highest $S1_{\textrm{spike}}$ bins are shown in Fig.~\ref{fig:nr_band_luxsim_spike_count_phd_representative_bin_sim_comparison}.
The middle and high-energy bins are well fit using a Gaussian, but non-Gaussian behavior is observed at low $S1_{\textrm{spike}}$.
This non-Gaussian behavior is expected due to the low number of signal carriers produced at the interaction site as well as the effect of the 164~phd $S2$ threshold. 
The simulated distribution of events in each nuclear recoil band bin is represented by the red shaded histogram in Fig.~\ref{fig:nr_band_luxsim_spike_count_phd_representative_bin_sim_comparison}.
The LUXSim simulation captures the non-Gaussian behavior at low-$S1$ and provides an accurate model of the nuclear recoil band in the profile likelihood ratio analysis used for the WIMP search results~\cite{AkeribAraujoBaiEtAl2015, AkeribAraujoBaiEtAl2016}.

\section{NEST model fit to D-D data} \label{sec:nest_post_dd}

To directly use the \qy{} and \ly{} measurements in LUX simulation and analysis, we performed a fit of the NEST model to the data presented in this paper.
We used a Metropolis-Hastings algorithm to sample a global likelihood function, in which the model was simultaneously constrained by the measurements of the nuclear recoil band mean (Section~\ref{sec:dd_nr_band}), light yield (Sections~\ref{sec:dd_low_energy_ly}~\&~\ref{sec:dd_endpoint}), and charge yield (Section~\ref{sec:dd_low_energy_qy}~\&~\ref{sec:dd_endpoint}).
The procedure followed the methodology described in Ref.~\cite{LenardoKazkazManalaysayEtAl2015}. 
The model parameterization and optimization are described in detail in Ref.~\cite{Akeribothers2016}, and the resultant NEST model is used in the analyses presented in Refs.~\cite{AkeribAraujoBaiEtAl2015, AkeribAraujoBaiEtAl2016}.
Below we discuss the implications for the physics of liquid xenon response at low energies.

In contrast to electronic recoils, recoiling nuclei lose a fraction of their energy to nuclear collisions, dissipating energy as heat rather than in processes leading to a detectable electronic signal.
Reconstruction of nuclear recoil event energy, therefore, requires an understanding of these processes as a function of recoil energy.
The formula for energy reconstruction can be written as

\begin{equation}
    E_{\textrm{nr}} = \frac{ W (N_{e} + N_{\textrm{ph}}) }{L} \,\text{,}
    \label{eq:EnergyScale}
\end{equation}

\noindent
where $L$ is the fraction of energy that goes into detectable electronic channels~\cite{Sorensen2011}.
Here, $W = 13.7~\text{eV}$ is the average energy needed to create an exciton or electron-ion pair, \Ne{} is the absolute number of ionization electrons, and \Nph{} is the absolute number of scintillation photons.
Both \Ne{} and \Nph{} represent the number of signal carriers after recombination but before biexcitonic quenching effects, in contrast to \nel{} and \np{} defined earlier in Secs.~\ref{sec:dd_low_energy_qy} and~\ref{sec:dd_low_energy_ly}, which are the measured number of signal carriers that escape the interaction site.
A detailed description of the recombination and biexcitonic quenching components of the model is reported in Ref.~\cite{Akeribothers2016}. 

The factor $L$ is traditionally given by the Lindhard model~\cite{Lindhard1963a, Sorensen2011}.
It is described by the formula 

\begin{equation}
    L = \frac{k \, g(\epsilon)}{1 + k\, g(\epsilon)} \,\text{.}
    \label{eq:Lindhard}
\end{equation}

\noindent
The parameter $k$ is a proportionality constant between the electronic stopping power and the velocity of the recoiling nucleus.
The quantity $g(\epsilon)$ is proportional to the ratio of electronic stopping power to nuclear stopping power, calculated using the Thomas-Fermi screening function.
It is a function of the energy deposited, converted to the dimensionless quantity $\epsilon$ using

\begin{equation}
    \epsilon = 11.5 (E_{\textrm{nr}} / \textrm{keV}_{\textrm{nr}}) Z^{-7/3} \,\text{.}
\end{equation}

\noindent
In these terms, $g(\epsilon)$ is given in Ref.~\cite{Lewin1996} by

\noindent
\begin{equation}
    g(\epsilon) = 3 \epsilon^{0.15} + 0.7 \epsilon^{0.6} + \epsilon \,\text{.}
\end{equation}

\noindent
A commonly accepted value for the proportionality constant is $k = 0.166$, but this may range from 0.1 to 0.2~\cite{Sorensen2011}.
We utilize the Lindhard model in our nuclear recoil response model, allowing $k$ to float in the fit to these data.
The best-fit value from the global optimization is $k = 0.1735 \pm 0.0060$.  

In addition to Lindhard's model, we explored an alternative model proposed in Ref.~\cite{Bezrukov2011} with a larger ionization and scintillation yield at recoil energies below 2~\kevnr{}.
To do so, we begin with the generic form of $L$ in Eq.~\ref{eq:EnergyScale}:

\begin{equation}
    L = \alpha \frac{s_{e}}{s_{e} + s_{n}} \,\text{.}
\end{equation}

\noindent
Here, \se{} and \sn{} are the electronic and nuclear stopping powers, respectively, and $\alpha$ is a scaling parameter used to model the cascade of collisions in a nuclear recoil event (best-fit is $\alpha = 2.31$ in the global optimization).
The ratio \se{}/\sn{} is analogous to $g$ in Eq.~\ref{eq:Lindhard}.
While the Lindhard model uses the Thomas-Fermi approximation to calculate \sn{}, we replace this with the empirical form from Ziegler et al.~\cite{ZieglerLittmarkBiersack1985}:

\begin{equation}
    s_n(\epsilon_\text{Z}) = \frac{\text{ln}(1 + 1.1383\,\epsilon_\text{Z})}{2(\epsilon_\text{Z} + 0.01321\,\epsilon_\text{Z}^{0.21226} + 0.19593\,\epsilon_\text{Z}^{0.5})} \,\text{,}
\end{equation}

\noindent
where $\epsilon_\text{Z} = 1.068\epsilon$.
The slight difference in energy scales is due to different assumed screening lengths in the calculation of the dimensionless energy. 

\begin{figure}[ht]
    \includegraphics[width=0.48\textwidth]{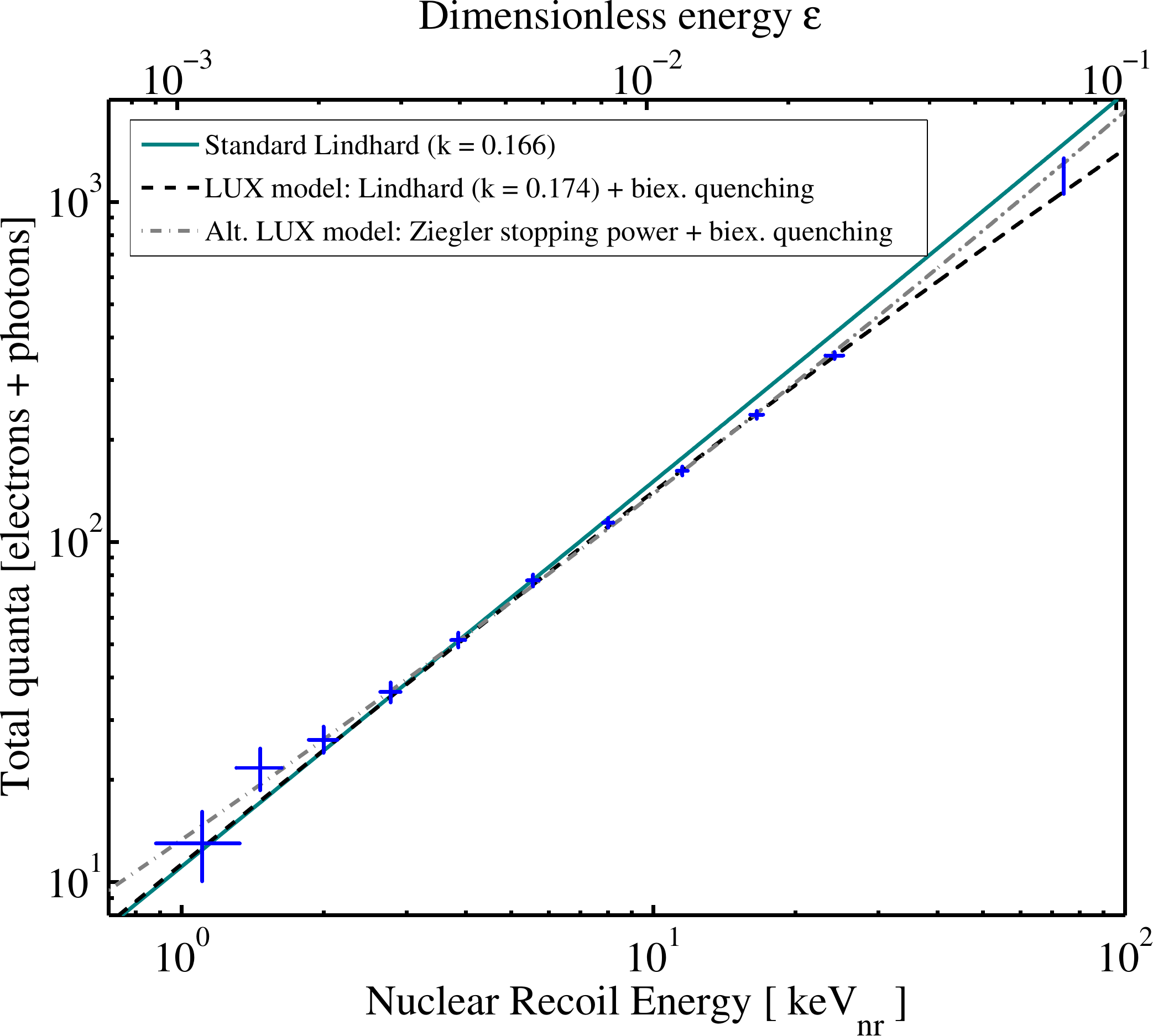}
    \caption{
        Total quanta, calculated by summing the measured light and charge yields.
        Predicted number of quanta using the two LUX nuclear recoil models described in this work and the standard
        Lindhard model are shown.
        The disagreement between the LUX models and the standard Lindhard model at high energies is due to our inclusion of biexcitonic interactions and Penning ionization.
        The net effect of these processes reduces the number of total quanta as the exciton density increases and better describes the data above 10~\kevnr{}.
    }
    \label{fig:nq_plot_for_dd_paper_draft}
\end{figure}

To directly compare to data, we sum the measured light and charge to get a measured total quanta, \nq{}~$=$~\nel{}~$+$~\np{}.
This is accomplished by interpolating the measured light yield using an empirical power law fit, and adding the result to the charge yield at the measured energies.
To avoid extrapolation of the light yield, we ignore the 0.70~\kevnr{} charge yield bin and consider only points above 1.08~\kevnr{}.
The fractional statistical uncertainties in light yield are also empirically interpolated and added in quadrature to the statistical uncertainties in \qy{} to estimate uncertainties in \nel{}.
The result is plotted against the total quanta predicted by our best-fit nuclear recoil models and the standard Lindhard model in Fig.~\ref{fig:nq_plot_for_dd_paper_draft}.
We find excellent agreement with the unmodified Lindhard model in the low energy regime down to 1.1~\kevnr{}.  

The disagreement with the Lindhard model at high energies ($>$10~\kevnr{}) is attributed to biexcitonic effects, in which two excitons can interact to produce only one photon, or one photon and one electron (Penning ionization).
Evidence for such effects in other experiments has been described in Refs.~\cite{Manzur2010, CaoAlexanderAprahamianEtAl2015}.
We incorporate this into our model via the quenching factor

\begin{equation}
    f_l = \frac{1}{1 + \eta \, s_{e}} \,\text{,}
\end{equation}

\noindent
where $s_{e} = 0.166 \, \epsilon ^{1/2}$ is the theoretical electronic stopping power for liquid xenon~\cite{Bezrukov2011} and $\eta$ is a free parameter allowed to float in the fit.
This factor multiplies the total number of predicted photons, and a fraction of this is added to the total number of predicted ionization electrons to model Penning ionization.
The optimal value obtained is $\eta = 13.2 \pm 2.3$.
The fraction of biexcitonic collisions resulting in ionization is modeled as an additional free parameter.
The inclusion of these effects allows our models to describe the data across the energy range spanned by \dd{} neutron induced recoils.
A more detailed discussion of both models, including a table of all best-fit parameters, is reported in Ref.~\cite{Akeribothers2016}. 

The model using the Ziegler stopping power is found to be a better description of our data below 2~\kevnr{}; however, it provides a slightly worse fit over the entire energy range (1--74~\kevnr{}).
Therefore, we employ the Lindhard-based (with $k = 0.1735$) NEST model in LUX data analysis and simulation.
As it is fit directly to the \insitu{} calibration data, this model produces a robust description of liquid xenon response for the simulation and reconstruction of nuclear recoil events within the LUX detector.

\section{Summary} \label{sec:lux_dd_summary}

A \dd{} source was used to produce a collimated beam of mono-energetic 2.45~MeV neutrons incident on the LUX detector.
This neutron source was used to characterize the nuclear recoil response of LUX \insitu{} in the dark matter detector itself.

The low-energy ionization yield result described in Sec.~\ref{sec:dd_low_energy_qy} was obtained using a new technique to directly measure the nuclear recoil energy using the reconstructed angle between interactions in double-scatter events in the LUX TPC.
The reported ionization yield has been measured a factor of $\times$5 lower in energy than any other previous calibration with a kinematically-defined energy scale.
The low-energy scintillation yield was measured using the single-scatter event population as described in Sec.~\ref{sec:dd_low_energy_ly}.
The reported scintillation yield has been measured a factor of $\times$3 lower in energy than has been achieved previously, and is the first liquid xenon \ly{} result reported in the absolute units of photons/\kevnr{}.
The resulting light and charge yields are consistent with other recent measurements in the literature, as shown in Fig.~\ref{fig:ionization_yield_qy_endpoint_ionization_yield} and Fig.~\ref{fig:scintillation_yield_alternate_alpha_scaling_s1_only_light_yield_final}.

In addition, the kinematically fixed 74~\kevnr{} endpoint of the nuclear recoil energy spectrum in liquid xenon was used to extract the charge and light yields as reported in Sec.~\ref{sec:dd_endpoint}.
The measured signal yields at the recoil spectrum endpoint are also consistent with previously reported results in the literature at similar recoil energy.

The ratio of ionization to scintillation, commonly used in liquid xenon TPCs to discriminate between nuclear and electron recoils, was measured for nuclear recoils in Sec.~\ref{sec:dd_nr_band}.
The collimated beam of neutrons from the \dd{} source provides a nuclear recoil band calibration with minimal contamination from multiple scintillation, single ionization events.
All nuclear recoil measurements were performed at an electric field of 180~V/cm.

After the measurements, two new versions of the NEST model were created using the simultaneous constraints provided by the measured \qy{}, \ly{}, and nuclear recoil band results as described in Sec.~\ref{sec:nest_post_dd}.
The first, more conservative, parameterization used for the recent LUX WIMP search results~\cite{AkeribAraujoBaiEtAl2015, AkeribAraujoBaiEtAl2016} was based upon the Lindhard model.
An alternative parameterization was based upon the Bezrukov model using the Ziegler stopping power.
Both the Lindhard and Bezrukov based models agree with the measured signal yields within experimental uncertainties over the entire two order of magnitude recoil energy range for which results are reported.

These results define the nuclear recoil signal response in both channels (charge and light) from 1.1 to 74~\kevnr{}, which covers the entire recoil energy range used for the LUX WIMP search.
The demonstration of signal yield in liquid xenon at recoil energies as low as 1.1~\kevnr{} provides an improved calibration of LUX sensitivity to low mass WIMPs---a factor of $\times$7 improvement in sensitivity for WIMPs of mass 7~GeV\,c$^{-2}$.
As a direct result of this calibration, the lowest kinematically accessible WIMP mass has been reduced from 5.2~to~3.3~GeV\,c$^{-2}$~\cite{AkeribAraujoBaiEtAl2015}.

This newly demonstrated nuclear recoil signal response below 3~\kevnr{} also enables improved estimates of expected coherent neutrino-nucleus scattering event rates in liquid xenon TPCs.
The recent LUX WIMP search results had the expectation of observing 0.10 such events due to $^{8}$B solar neutrinos under the LUX Lindhard model, while the Bezrukov model provides an expectation of 0.16 observed events~\cite{AkeribAraujoBaiEtAl2015}.

\begin{acknowledgments}

    This work was partially supported by the U.S. Department of Energy (DOE) under award numbers DE-FG02-08ER41549, DE-FG02-91ER40688, DE-FG02-95ER40917, DE-FG02-91ER40674, DE-NA0000979, DE-FG02-11ER41738, DE-SC0006605, DE-AC02-05CH11231, DE-AC52-07NA27344, DE-FG01-91ER40618, and DE-SC0010010; the U.S. National Science Foundation under award numbers PHYS-0750671, PHY-0801536, PHY-1004661, PHY-1102470, PHY-1003660, PHY-1312561, PHY-1347449, PHY-1505868, PHY-1636738, and PHY-0919261; the Research Corporation grant RA0350; the Center for Ultra-low Background Experiments in the Dakotas (CUBED); and the South Dakota School of Mines and Technology (SDSMT).
    LIP-Coimbra acknowledges funding from Funda\c{c}\~{a}o para a Ci\^{e}ncia e Tecnologia (FCT) through the project-grant PTDC/FIS-NUC/1525/2014.
    Imperial College and Brown University thank the UK Royal Society for travel funds under the International Exchange Scheme (IE120804).
    The UK groups acknowledge institutional support from Imperial College London, University College London and Edinburgh University, and from the Science \& Technology Facilities Council for PhD studentships ST/K502042/1 (AB), ST/K502406/1 (SS) and ST/M503538/1 (KY).
    The University of Edinburgh is a charitable body, registered in Scotland, with registration number SC005336.

    This research was conducted using computational resources and services at the Center for Computation and Visualization, Brown University.
    We thank Alan Robinson for providing useful feedback on a preprint of this manuscript.

    We gratefully acknowledge the logistical and technical support and the access to laboratory infrastructure provided to us by the Sanford Underground Research Facility (SURF) and its personnel at Lead, South Dakota.
    SURF was developed by the South Dakota Science and Technology Authority, with an important philanthropic donation from T. Denny Sanford, and is operated by Lawrence Berkeley National Laboratory for the Department of Energy, Office of High Energy Physics.

\end{acknowledgments}

\bibliographystyle{apsrev4-1}

\bibliography{lux_run03_dd_prc}

\end{document}

%% file: lux_run03_dd_author_list.tex
\author{D.S.~Akerib} %
\affiliation{Case Western Reserve University, Department of Physics, 10900 Euclid Ave, Cleveland, OH 44106, USA}
\affiliation{SLAC National Accelerator Laboratory, 2575 Sand Hill Road, Menlo Park, CA 94205, USA}
\affiliation{Kavli Institute for Particle Astrophysics and Cosmology, Stanford University, 452 Lomita Mall, Stanford, CA 94309, USA}

\author{S.~Alsum} %
\affiliation{University of Wisconsin-Madison, Department of Physics, 1150 University Ave., Madison, WI 53706, USA}

\author{H.M.~Ara\'{u}jo} %
\affiliation{Imperial College London, High Energy Physics, Blackett Laboratory, London SW7 2BZ, United Kingdom}

\author{X.~Bai} %
\affiliation{South Dakota School of Mines and Technology, 501 East St Joseph St., Rapid City, SD 57701, USA}

\author{A.J.~Bailey} %
\affiliation{Imperial College London, High Energy Physics, Blackett Laboratory, London SW7 2BZ, United Kingdom}

\author{J.~Balajthy} %
\affiliation{University of Maryland, Department of Physics, College Park, MD 20742, USA}

\author{P.~Beltrame} %
\affiliation{SUPA, School of Physics and Astronomy, University of Edinburgh, Edinburgh EH9 3FD, United Kingdom}

\author{E.P.~Bernard} %
\affiliation{University of California Berkeley, Department of Physics, Berkeley, CA 94720, USA}
\affiliation{Yale University, Department of Physics, 217 Prospect St., New Haven, CT 06511, USA}

\author{A.~Bernstein} %
\affiliation{Lawrence Livermore National Laboratory, 7000 East Ave., Livermore, CA 94551, USA}

\author{T.P.~Biesiadzinski} %
\affiliation{Case Western Reserve University, Department of Physics, 10900 Euclid Ave, Cleveland, OH 44106, USA}
\affiliation{SLAC National Accelerator Laboratory, 2575 Sand Hill Road, Menlo Park, CA 94205, USA}
\affiliation{Kavli Institute for Particle Astrophysics and Cosmology, Stanford University, 452 Lomita Mall, Stanford, CA 94309, USA}

% ...
